\documentclass[%
 aip,
 amsmath,amssymb,
 reprint,%
]{revtex4-1}

\usepackage{nomencl}

\usepackage{graphicx}
\usepackage{dcolumn}
\usepackage{bm}

\usepackage[utf8]{inputenc}
\usepackage[T1]{fontenc}
\usepackage{mathptmx}
\usepackage{etoolbox}

\makeatletter
\def\@email#1#2{%
 \endgroup
 \patchcmd{\titleblock@produce}
  {\frontmatter@RRAPformat}
  {\frontmatter@RRAPformat{\produce@RRAP{*#1\href{mailto:#2}{#2}}}\frontmatter@RRAPformat}
  {}{}
}%
\makeatother
\begin{document}

\preprint{AIP/123-QED}

\title[Programmable Microwave Cluster States via Josephson Metamaterials]{Programmable Microwave Cluster States via Josephson Metamaterials}

\author{A. Alocco}
 \altaffiliation[Also at ]{Quantum Metrology and Nanotechnology Division, Istituto Nazionale di Ricerca Metrologica, Strada delle Cacce 91, Turin 10135, Italy}
\author{A. Celotto}%
 \altaffiliation[Also at ]{Quantum Metrology and Nanotechnology Division, Istituto Nazionale di Ricerca Metrologica, Strada delle Cacce 91, Turin 10135, Italy}
\author{E. Palumbo}%
 \altaffiliation[Also at ]{Quantum Metrology and Nanotechnology Division, Istituto Nazionale di Ricerca Metrologica, Strada delle Cacce 91, Turin 10135, Italy}
\affiliation{ 
Department of Applied Science and Technology, Politecnico di Torino, C.So Duca Degli Abruzzi 24, Turin 10129, Italy
}%

\author{B. Galvano}
\author{P. Livreri}
\affiliation{%
Department of Engineering, University of Palermo, Palermo, Italy
}%

\author{L. Fasolo}%
\author{L. Callegaro}%
\author{E. Enrico}%
 \email{e.enrico@inrim.it}
\affiliation{ 
Quantum Metrology and Nanotechnology Division, Istituto Nazionale di Ricerca Metrologica, Strada delle Cacce 91, Turin 10135, Italy
}%

\date{\today}

\begin{abstract}
Cluster states are a fundamental resource for continuous-variable quantum computing, enabling measurement-based protocols that can scale beyond the limitations of qubit-based architectures. Here, we demonstrate on-demand generation of multimode entangled microwave cluster states using a programmable Josephson Traveling-Wave Parametric Amplifier (JTWPA) operated in the three-wave mixing regime. By injecting a tailored, non-equidistant set of pump tones via an arbitrary waveform generator, we engineer frequency-specific nonlinear couplings between multiple frequency modes. The entanglement structure is verified via frequency-resolved heterodyne detection of quadrature nullifiers, confirming the target graph topology of the cluster state. Our approach allows reconfigurability through the pumps spectrum and supports scalability by leveraging the wide bandwidth and spatial homogeneity of the JTWPA. This platform opens new avenues for scalable measurement-based quantum information processing in the microwave domain, compatible with superconducting circuit architectures.
\end{abstract}

\maketitle

\section{\label{sec:Theory}Introduction}
Quantum information processing with continuous variables (CV) provides a powerful framework for realizing scalable quantum computation, built upon Gaussian states, deterministic operations, and high-efficiency measurements \cite{Braunstein2005,Weedbrook2012}. A central resource in this paradigm is the CV cluster state, a highly entangled multipartite state that underpins measurement-based quantum computing (MBQC)~\cite{Zhang2006,Menicucci2006,Yokoyama2013}.
In the optical domain, the generation of large-scale cluster states has been successfully demonstrated using time-multiplexed or frequency-encoded modes~\cite{Yokoyama2013,Asavanant2019,Larsen2019,Jia2025}. These implementations typically rely on dielectric nonlinearities and optical interferometry~\cite{Reimer2016}. However, transitioning CV cluster state generation to the microwave regime holds great promise, particularly in the context of superconducting circuits. The microwave platform offers compatibility with circuit quantum electrodynamics (cQED), coherent interaction mechanisms, and the potential for fully integrated and reconfigurable quantum processors~\cite{Gu2017,Blais2021}.

Entanglement of microwave modes has been demonstrated with Josephson parametric amplifiers (JPAs) and Josephson mixers~\cite{Eichler2011,Flurin2012}, including the realization of two-mode squeezing and multipartite configurations~\cite{Pfaff2017,SandboChang2018,Petrovnin2023,Lingua2025}. Nevertheless, these approaches typically involve discrete resonators or fixed topologies, which limit the scalability and flexibility needed for large-scale MBQC.

In this work, we present a new platform based on a Josephson Traveling-Wave Parametric Amplifier (JTWPA)~\cite{macklin2015near} operating in the three-wave mixing regime, which enables the deterministic generation of reconfigurable and scalable cluster states across multiple microwave modes. By driving the JTWPA with a custom-designed set of non-equidistant pump tones, we selectively activate mode couplings corresponding to the desired graph connectivity. The use of broadband traveling-wave interaction allows simultaneous generation of multiple entangled pairs, forming extended cluster graphs.

We characterize the resulting quantum correlations by heterodyne detection of frequency-resolved nullifier observables, which serve as witnesses for the underlying entanglement structure~\cite{Menicucci2011,Cai2017}. The flexibility of our platform is illustrated by tailoring various topologies via the pump spectral profile. We further discuss how this architecture lends itself to future integration with quantum feedback, error correction, and hybrid quantum systems.\\

\begin{figure*}
\includegraphics[width=\textwidth]{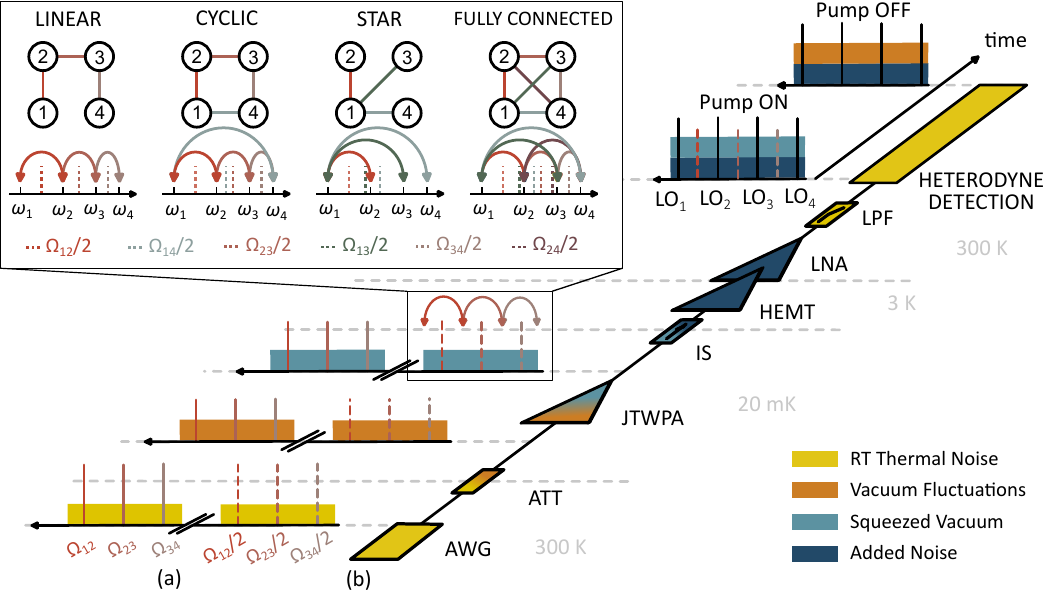}
\caption{\label{fig:1} \textbf{Programmable generation of microwave cluster states using a Josephson Traveling-Wave Parametric Amplifier.}  
(a) Schematic representation of the frequency-domain continuous-variable (CV) cluster state protocol. Multiple non-equidistant pump tones, synthesized via an Arbitrary Waveform Generator (AWG), drive a JTWPA into a three-wave mixing regime, realizing entangling interactions on the vacuum fluctuations.  
(b) Block diagram of the experimental setup, featuring microwave multi-mode synthesis, cryogenic routing, and multiplexed heterodyne detection using a reference-state protocol for rejecting added readout line noise (pumps ON/OFF). Inset: various cluster topologies and their associated pump configurations are shown. Each edge between modes \( (j, l) \) is implemented by applying a pump tone at frequency \( \Omega_{jl} = \omega_j + \omega_l \).}

\end{figure*}

\section{Programmable Generation of Cluster States}

Continuous-variable (CV) cluster states are defined by a graph structure where each node represents a bosonic mode and edges correspond to entangling interactions~\cite{Menicucci2006, Zhang2006}. In the frequency domain, this structure can be implemented by coupling different frequency modes via nonlinear mixing processes~\cite{Roslund2014, Ferrini2013}.

We implement this coupling using a JTWPA operating in the three-wave mixing regime. The effective interaction is described by the Hamiltonian
\begin{equation}
\label{eq:squeezparam}
    \hat{H}_\mathrm{eff} = \sum_{k=1}^n \sum_{j \le l} \left( r_{jl}^{(k)} \hat{a}_j^\dagger \hat{a}_l^\dagger + r_{jl}^{(k)*} \hat{a}_j \hat{a}_l \right),
\end{equation}
where $\hat{a}_j$ and $\hat{a}_l$ are the annihilation operators for modes $j$ and $l$, with respective frequencies $\omega_j$ and $\omega_l$. The parameter $r_{jl}^{(k)}$ denotes the squeezing amplitude associated with the $k$-th pump tone of frequency $\Omega_{jl}^{(k)}$, which satisfies the energy conservation condition $\omega_j + \omega_l = \Omega_{jl}^{(k)}$. The total number of pump tones, $n$, is determined by the chosen coupling topology (see inset in Figure~\ref{fig:1}).

To enable flexible control over the interaction graph, we synthesize a programmable set of pump tones using an arbitrary waveform generator (AWG). These pumps span the JTWPA’s operational bandwidth and are non-equidistant to match desired pairwise resonances. This method enables arbitrary graph structures among a selected set of frequency modes, akin to an optical frequency comb~\cite{Yokoyama2013, Asavanant2019, Larsen2019} (scheme reported in Figure \ref{fig:1} for a four-mode linear cluster).

Compared to conventional JPA-based architectures, our approach eliminates the need for cavity storage or delay lines, enabling entanglement over several GHz bandwidth. The programmable nature of the AWG-based pumps allows real-time reconfiguration of the graph structure, making the system well-suited for dynamic CV quantum protocols~\cite{Chen2014}.

Cluster state generation is verified using multiplexed heterodyne or homodyne detection and nullifier measurements~\cite{Cai2017, Jia2025, Lingua2025}, allowing reconstruction of the graph structure from measured correlations (see Figure \ref{fig:1}). We find high-fidelity agreement with the targeted graph structure, demonstrating the viability of our method for scalable, reconfigurable CV quantum networks.

Further details on the Hamiltonian model and pump-tone selection strategy can be found in the Supplementary Information.

\begin{figure*}
\includegraphics[width=\textwidth]{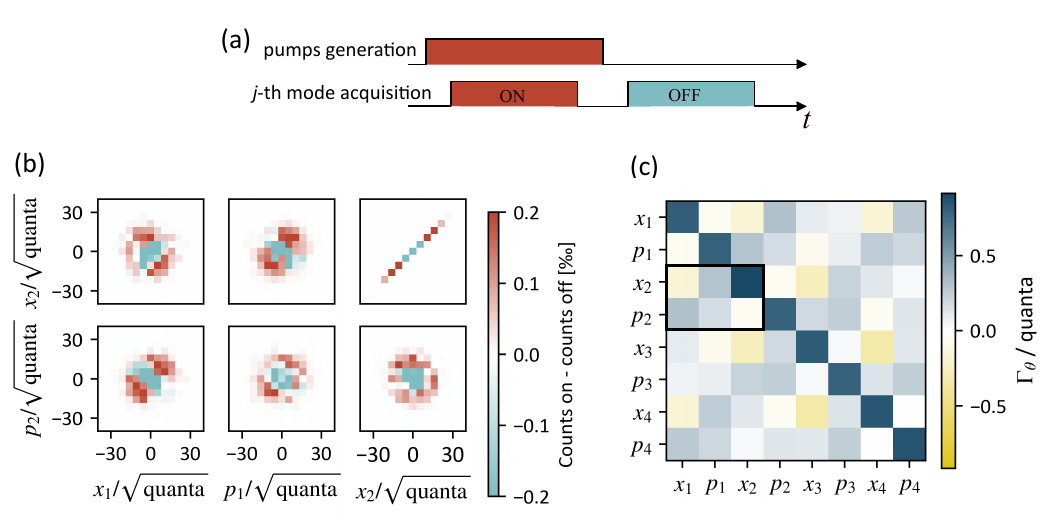}
\caption{\label{fig:2} \textbf{Gaussian cluster state covariance matrix reconstruction.}
(a) Scheme of the pumps interleaved generation via an ON-OFF protocol, which guarantees referenced-state multiplexed heterodyne detection of modes. 
(b) Histograms reporting the counts of quadratures measured in the ON condition, subtracted using the reference state acquired during the OFF state for modes 1 and 2. Correlations are reported in the pairs $(x_2$,$p_1)$ and $(p_2,x_1)$, while anti-correlations are in $(x_2,x_1)$.   
(c) Covariance matrix in the yet unknown $\theta$-rotated basis $ \Gamma_\theta$ reconstructed from multiplexed heterodyne measurements on $4$ modes cyclically clustered via $4$ pumps. The black box indicates the elements of the covariance matrix corresponding to the data reported in (b).}
\end{figure*}

\section{Measurement and Nullifier Verification}

To verify the presence of multipartite entanglement and validate the underlying graph structure of the generated cluster state, we employ nullifier-based measurements~\cite{Menicucci2011, Cai2017}. In the continuous-variable model, an ideal cluster state associated with an adjacency matrix $\mathbf{A}$ satisfies a set of operator equations called nullifiers:
\begin{equation}
    \hat{\delta}_j = \hat{p}_j - \sum_l A_{jl} \hat{x}_l \rightarrow 0,
\end{equation}
where $\hat{x}_l$ and $\hat{p}_j$ are the amplitude and phase quadratures, respectively, of modes $l$ and $j$, expressed in the photon basis, and $A_{jl}$ is the real-valued weight of the connection between modes $j$ and $l$.

In practice, finite squeezing leads to non-zero nullifier variances. A state is considered a valid cluster state if all nullifier variances are suppressed below the shot noise level (SNL), also known as the quantum limit~\cite{Gu2009, Adesso2014}. We perform heterodyne detection on each frequency mode, properly tuning a set of Local Oscillators (LO) and exploiting a digital down-conversion scheme to extract the correlations between different quadratures over time windows synchronized with the pump tone cycle.

\begin{figure*}
\includegraphics[width=\textwidth]{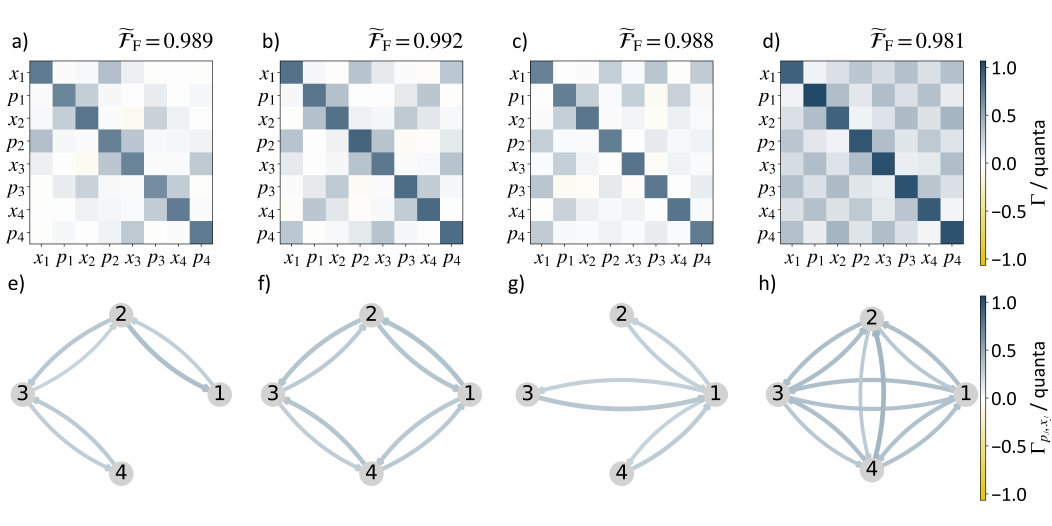}
\caption{\label{fig:3} \textbf{Experimental validation of cluster graph and correlations via adjacency matrix reconstruction and nullifier measurements.}  
(a-d) Reconstructed covariance matrix $ \Gamma$ of the quadrature operators, extracted from heterodyne detection of four frequency modes clustered in linear (a,e), cyclic (b,f), star (c,g), and fully-connected (d,h) configurations.  
(e-f) Evaluation of the graph connectivity inferred from the  elements of the covariance matrix, with the colour of the arrow pointing from the \textit{j}-th to the \textit{l}-th node mapping the value of the $\Gamma_{p_j,x_l}$ term.}
\end{figure*}

From the reconstructed covariance matrix in the yet unknown $\theta$-rotated basis $ \Gamma_\theta$ (see Figure \ref{fig:2}), we recover the cluster state covariance matrix $\Gamma$ via numerical optimization of local transformations (see Figure \ref{fig:3}(a-d)), maximizing the off-diagonal $x,p$ correlations. The corresponding network graphs confirm the programmed topology and measurement-based reconstruction of the entangled state (see Figure \ref{fig:3}(e-h)). The experimental matrices are then compared via the Frobenius distance ($\widetilde{{\mathcal{{F}}}}_\text{{F}}$) with theoretical predictions for multipartite entanglement~\cite{vanLoock2003}.
The worst-case Frobenius distance deviates by less than 2\% from the theoretical expectation, demonstrating quantitative consistency with the target cluster state.

\begin{figure}
\includegraphics[width=\linewidth]{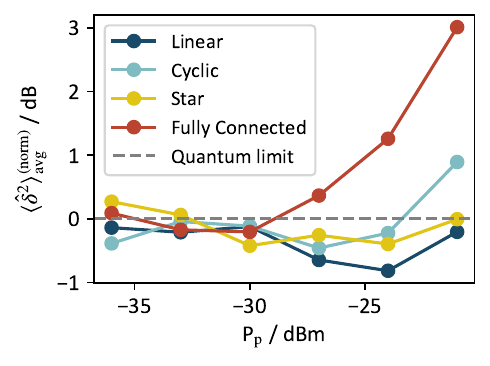}
\caption{\label{fig:4} \textbf{Cluster quality metrics for different topologies and pump powers.} Dependence of the average nullifier variance, normalized to the shot noise level, \( \langle \hat{\delta}^2 \rangle_{\mathrm{avg}}^{\text{(norm)}} \), on the squeezing amplitude of the mode pairs, controlled by \( P_\text{p} \), the power of a single pump tone (identical for all pairs) measured at the room-temperature reference plane, for different cluster topologies. This behavior highlights the dynamic reconfigurability of the entanglement graph.
}
\end{figure}

Our results, reported in Figure \ref{fig:4}, show that all nullifier variances fall below the quantum limit for $N=4$ modes, in a wide range of pump powers, for four different graph topologies depending on the pump configuration.

The full derivation of the nullifiers, the construction of the adjacency matrix, and numerical reconstruction techniques are provided in the Supplementary Information.

\section{Experimental Setup}

The experimental platform is built around a pre-commercial Al-TWPA-C from Arctic Instruments, consisting of a periodic array of Josephson junction-based meta-atoms embedded in a thin-film waveguide structure \cite{Perelshtein2022}. The JTWPA supports broadband non-degenerate three-wave mixing when biased with an external DC flux and driven by a programmable set of pump tones.

Pump tones are synthesized using a high-speed arbitrary waveform generator (AWG) and digitally I/Q-modulated onto a microwave carrier before attenuation and injection into the JTWPA input port.

The broadband vacuum fluctuations act as the quantum seed for spontaneous two-mode squeezing via parametric interactions. The output of the JTWPA is passed through an isolator, low-noise amplifiers, and a room-temperature down-conversion chain before being digitized by a fast analog-to-digital converter (ADC). The acquired signals are processed using digital signal processing (DSP) to perform mode demultiplexing and quadrature extraction~\cite{Pfaff2017, Krantz2019}.

The entire setup is housed in a cryogen-free dilution refrigerator, which operates below 20 mK to minimize thermal noise. Signal integrity is preserved using superconducting coaxial cables, attenuators, and infrared filters at proper temperature stages. Phase coherence between pump, seed, and LO paths is actively stabilized using a reference tone distributed throughout the setup.

Calibration of squeezing levels, phase drifts, and JTWPA efficiencies is performed using cryogenic standard references.

A comprehensive description of the cryogenic setup, signal routing, and DSP implementation is included in the Supplementary Information.

\section{Discussion and Outlook}

Our demonstration of programmable continuous-variable (CV) cluster state generation using a broadband Josephson parametric amplifier represents a significant advance toward scalable and reconfigurable photonic quantum architectures. By encoding quantum modes in the frequency domain and engineering entanglement via pump tone synthesis, we eliminate the need for bulky optical elements or time-multiplexed delay lines traditionally used in CV platforms~\cite{Asavanant2019, Larsen2019, Chen2014}.
The programmability of our approach enables rapid reconfiguration of the interaction graph, a capability critical for implementing measurement-based quantum algorithms and quantum error correction in the CV regime~\cite{Menicucci2014, Fukui2018, Bourassa2021}. Combined with high detection efficiency and verified suppression of the nullifier variances across multiple frequency modes, this establishes a practical route toward real-time continuous-variable quantum processing using integrated superconducting hardware.

Current limitations include the finite squeezing generation~\cite{grimsmo2017squeezing}, the use of a HEMT amplifier at 3 K as the first amplification stage, room-temperature pump tones filtering, and residual four-wave mixing processes. These limitations can be mitigated, respectively, by employing a left-handed JTWPA \cite{kamal2024}, using a JTWPA as the first-stage amplifier, implementing cryogenic pump filtering, and optimizing the JTWPA operating point through dedicated flux tuning.

Furthermore, incorporating feedback mechanisms and photon-number-resolving detection may enable fault-tolerant CV quantum operations~\cite{Noh2020, Takeda2017}.

Looking ahead, our architecture is naturally extensible to higher-order graphs and frequency-mode multiplexing with thousands of channels, compatible with ongoing developments in cryogenic AWGs and integrated superconducting photonics~\cite{Zhong2020, Arute2019}. The ability to perform adaptive measurements and conditional feedforward using FPGA-based control further opens the path toward full-scale measurement-based quantum computation.
Our work thus bridges the gap between optical CV platforms and microwave quantum hardware, offering a flexible and hardware-efficient approach for continuous-variable quantum information processing.

\section{Conclusion}

In summary, we demonstrated a reconfigurable and scalable platform for continuous-variable cluster state generation based on programmable pump engineering in a superconducting Josephson parametric amplifier. Our approach leverages the frequency domain to synthesize large-scale entangled graphs with minimal overhead, verified through nullifier measurements and graph reconstruction. By bridging optical CV protocols with microwave superconducting hardware, this work opens a new route toward universal, hardware-efficient measurement-based quantum computation. Future integration with fast feedback may enable the real-time execution of quantum algorithms, advancing both foundational research and practical applications in quantum simulation and information processing.

\section*{Acknowledgments}
We would like to thank the team at VTT Technical Research Centre of Finland Ltd for providing the JTWPA: Debopam Datta, Wisa Förbom, Joonas Govenius, Robab Najafi Jabdaraghi, Janne Lehtinen, Jaani Nissilä, Mika Prunnila, Jorden Senior, Nils Tiencken, and Visa Vesterinen.\\
This work is partially supported by the European project MiSS. MiSS is funded by the European Union through the Horizon Europe 2021-2027 Framework Programme, Grant agreement ID: 101135868. This work is partially supported by the European project MetSuperQ. The 23FUN08 MetSuperQ project has received funding from the European Partnership on Metrology, co-financed from the European Union’s Horizon Europe Research and Innovation Programme and by the Participating States. This work is partially supported by the Italian project CalQuStates. The PRIN 2022 CalQuStates project received funding by the European Union – Next Generation EU Mission 4 Component 1 CUP E53D23002210006

\section*{Authors' Contribution}
A.A., A.C., B.G., E.E., and L.C. contributed to the conceptualization and design of the study. A.A., A.C., B.G., E.E., and L.F. performed the experiments. A.A., A.C., B.G., and E.E. developed the software, curated the data, and performed the analysis and validation. Methodology and visualization were contributed by A.A., A.C., B.G., E.E., E.P., L.F., and L.C. Writing of the manuscript involved A.A., A.C., B.G., E.E., E.P., L.F., L.C., and P.L. Supervision and project administration were provided by E.E., L.C., and P.L., who also secured funding and resources. All authors contributed to the experimental setup and manuscript preparation.

\section*{Bibliography}

\newpage

\begin{widetext}

\newpage
\thispagestyle{empty}
\mbox{}
\newpage

\renewcommand{\eqdeclaration}[1]{. See equation~(#1)}
\renewcommand{\pagedeclaration}[1]{, p.~#1.}
\makenomenclature
\newpage

\begin{center}
  {\LARGE\bfseries\rmfamily
  Supplementary Information:\\[0.6em]
  Programmable Microwave Cluster States via Josephson Metamaterials}
\end{center}

\vspace{2em}

\printnomenclature

\newpage

\setcounter{section}{0}
\setcounter{equation}{0}
\setcounter{figure}{0}

\section{Detailed Experimental Scheme}

\begin{figure}[h!]
    \centering
    \includegraphics[width=0.75\linewidth]{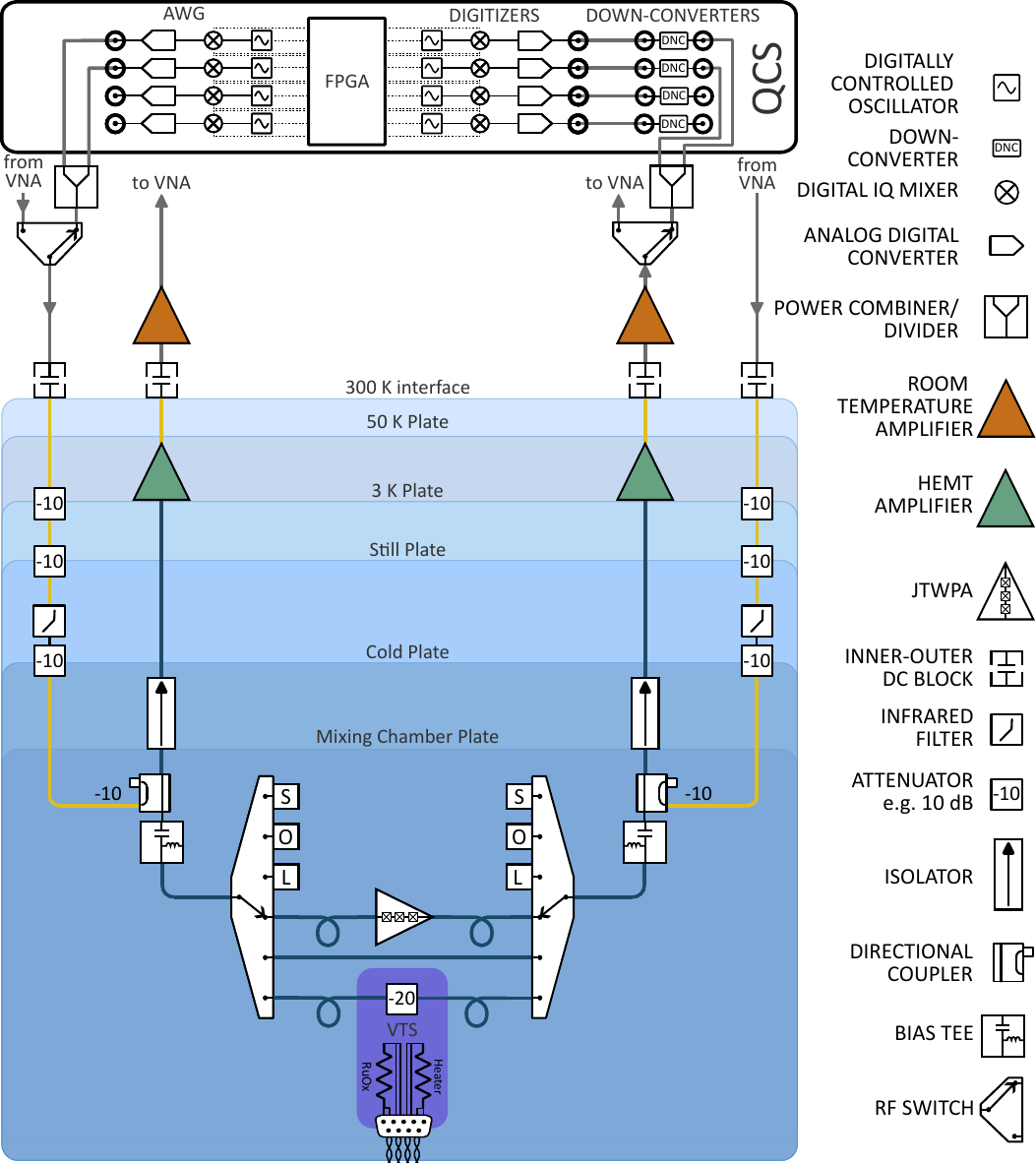} 
    \caption{\textbf{Schematic of the full cryogenic and room-temperature measurement setup.} The experiment is hosted in a dilution refrigerator, where two input lines are coupled to two output lines at the mixing chamber plate. Two cryogenic electromechanical switches allow for SOLR deembedding and in situ noise calibration using a variable temperature stage connected to a thermalized 20 dB attenuator. At room temperature, a Quantum Control System comprising AWG, FPGA, downconverters, and digitizers enables fully digital pump generation and heterodyne detection. Two additional switches allow routing to a nonlinear vector network analyzer (NVNA) for preliminary characterization \cite{Alocco25a}.}
    \label{fig:Setup}
\end{figure}

\newpage

\section{Pumps and Modes Schemes Selection and JTWPA Gain Flatness Validation}
\label{sec:2}

To ensure optimal entanglement conditions for the generation of continuous-variable cluster states, particular care was devoted to the positioning of the pump tones and frequency modes with respect to the gain profile of the Josephson Traveling Wave Parametric Amplifier (JTWPA).

\subsection{JTWPA Gain Flatness}

Figure~\ref{fig:JTWPA_Gain} displays the gain profiles of the JTWPA chip, embedded in its packaging and connected to electromechanical switches via uncompensated cables~\cite{oberto2025}, as measured across all pump frequencies used in this experiment.
In the frequency range of interest, centered around $4.5$ GHz, the amplifier provides a sufficiently flat gain profile for all the generated pump configurations. This behavior guarantees minimal amplitude distortion and phase dispersion, which are critical for preserving the entanglement structure of the generated cluster states.

\begin{figure}[h]
    \centering
    \includegraphics[width=\linewidth]{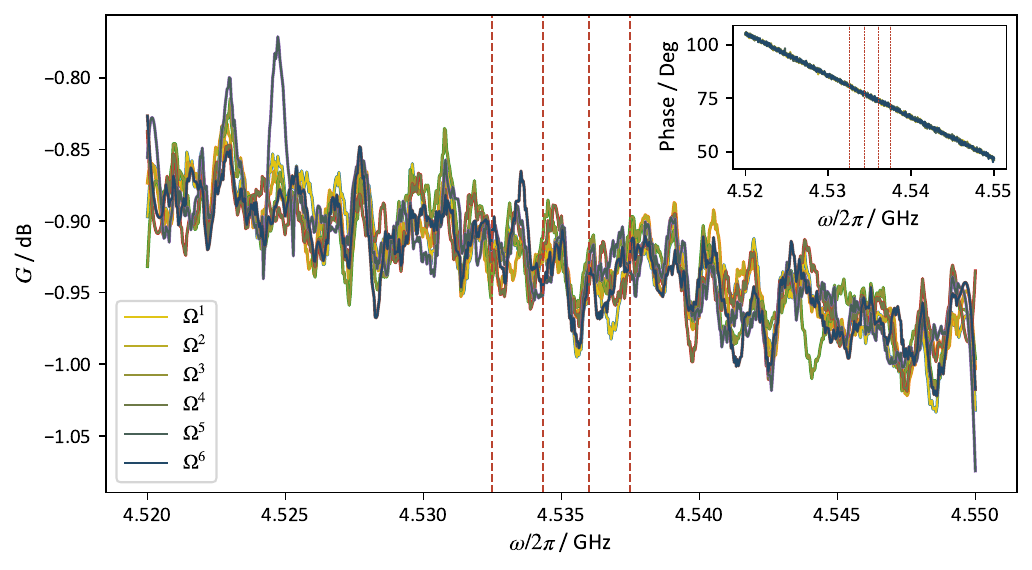}
    \caption{\textbf{Josephson Traveling Wave Parametric Amplifier Preliminary Characterization.} JTWPA gain $G$ as a function of signal frequency for six pump frequencies $\Omega^{(k)}$, at fixed pump power $P_{\Omega} = -21\text{ dBm}$. The amplifier is flux-biased at $\Phi = 0.378\text{ }\Phi_0$ to ensure predominant three-wave mixing. Dashed vertical lines indicate the modes of the investigated cluster. The inset shows the phase of the transmission coefficient $S_{21}$.}
    \label{fig:JTWPA_Gain}
\end{figure}



\subsection{Spectral Engineering for Cluster State Generation}

To study the impact of spectral arrangement on the distribution of entanglement in the generated cluster states, 
we engineered the mode frequency distribution using a sinusoidally modulated spacing:
\begin{equation}
    \omega_j = \omega_j + (j-1)\Delta\omega + A \sin\left(\frac{\pi (j)}{N-1}\right), \quad j = 1, 2, \ldots, N,
\end{equation}
\nomenclature{$\omega_j$}{Frequency of the $j$-th bosonic mode\nomrefeqpage}
where $\Delta\omega$ is a uniform base spacing, $A$ is the modulation amplitude, and $N$ is the total number of modes.

In the experimental implementation, the four mode frequencies were:
\begin{align}
    \omega_1 &= 4.5325 \times 10^9~\text{Hz}, \nonumber \\
    \omega_2 &= 4.5343\ldots \times 10^9~\text{Hz}, \nonumber \\
    \omega_3 &= 4.5360\ldots \times 10^9~\text{Hz}, \nonumber \\
    \omega_4 &= 4.5375 \times 10^9~\text{Hz}.
\end{align}

Pump tones were designed by computing all pairwise interactions using:
\begin{equation}
    \Omega^{(k)}_{jl} = \omega_j + \omega_l,
\end{equation}
for each pair $(j, l)$ defined by the target cluster topology.

The sinusoidal modulation plays a critical role in cluster state generation. 
Specifically, it prevents degeneracies in the pump tone frequencies which are used to establish pairwise entanglement. Without modulation, uniform mode spacing can cause multiple pairs to share identical $\Omega_{jl}$ values, leading to ambiguous interaction topologies and reduced control over the entanglement graph.
By introducing the sinusoidal term, each $\Omega_{jl}$ becomes distinct, enabling precise and unambiguous implementation of the desired cluster connectivity.

The supported entanglement graph configurations include:
\begin{itemize}
    \item \textbf{Linear:} Nearest-neighbor connections.
    \item \textbf{Star:} A central hub mode connected to all others.
    \item \textbf{Cyclic:} Ring topology with periodic boundary conditions.
    \item \textbf{Fully Connected:} All-to-all entanglement.
\end{itemize}

\subsection{Discussion}

The combination of flat JTWPA gain in the targeted frequency band and the systematic generation of pump-mode configurations allows us to synthesize a variety of cluster-state topologies. These results validate the suitability of the amplifier for scalable continuous-variable quantum computing platforms.

\newpage

\section{Theoretical Framework}

\subsection{Introduction}

Nonlinear media driven by multiple coherent pump tones enable the generation of multimode entangled states via parametric interactions. Such platforms are central to continuous-variable (CV) quantum information processing, and are particularly relevant for the generation of Gaussian cluster states used in measurement-based quantum computing. 

In this work, we examine the most general scenario of $m$ bosonic modes coupled through $n$ pump tones, each characterized by arbitrary amplitude, frequency, and phase. We investigate the resulting squeezing properties and mode correlations by analyzing the output covariance matrix and introducing the concept of a cluster nullifier.

\subsection{Model Hamiltonian}

We consider a bosonic system composed of \( m \) modes described by annihilation operators \( \hat{a}_j \) with free Hamiltonian
\begin{equation}
    \hat{H}_0 = \sum_{j=1}^m \omega_j \hat{a}_j^\dagger \hat{a}_j,
\end{equation}
\nomenclature{$m$}{Total number of bosonic modes\nomrefeqpage}
\nomenclature{$\hat{H}_0$}{Free Hamiltonian of the bosonic modes\nomrefeqpage}
\nomenclature{$\hat{a}_j$}{Annihilation operator for the $j$-th mode\nomrefeqpage}
\nomenclature{$\hat{a}_j^\dagger$}{Creation operator for the $j$-th mode\nomrefeqpage}
where \( \omega_j \) is the frequency of the \( j \)-th mode.
The system is driven by \( n \) classical pump tones, each described by frequency \( \Omega_{jl}^{(k)} \), amplitude \( \alpha_k \), and phase \( \phi_k \). These tones induce three-wave mixing interactions via a second-order nonlinearity.

The interaction Hamiltonian in the rotating wave approximation reads
\begin{equation}
    \hat{H}_\mathrm{int}(t) = \sum_{k=1}^n \sum_{j \le l} g^{(k)}_{jl} \alpha_k e^{-i \Omega_{jl}^{(k)} t - i\phi_k} \hat{a}_j^\dagger \hat{a}_l^\dagger + \mathrm{H.c.},
\end{equation}
\nomenclature{$n$}{Number of classical pump tones\nomrefeqpage}
\nomenclature{$\Omega_{jl}^{(k)}$}{Frequency of the $k$-th pump tone\nomrefeqpage}
\nomenclature{$g^{(k)}_{jl}$}{Nonlinear coupling coefficient between modes $j$ and $l$ under pump $k$\nomrefeqpage}
\nomenclature{$\hat{H}_\mathrm{int}(t)$}{Interaction Hamiltonian in the lab frame\nomrefeqpage}
where \( g^{(k)}_{jl} \in \mathbb{C} \) encodes the nonlinear coupling between modes \( j \) and \( l \) under pump \( k \). Symmetry \( g^{(k)}_{jl} = g^{(k)}_{lj} \) is assumed. The total Hamiltonian is
\begin{equation}
    \hat{H}(t) = \hat{H}_0 + \hat{H}_\mathrm{int}(t).
\end{equation}
\nomenclature{$\hat{H}(t)$}{Total Hamiltonian in the lab frame\nomrefeqpage}
In the interaction picture with respect to \( \hat{H}_0 \), the bosonic operators evolve as
\begin{equation}
    \hat{a}_j(t) = \hat{a}_j e^{-i \omega_j t},
\end{equation}
and the interaction Hamiltonian becomes
\begin{equation}
    \hat{H}_I(t) = \sum_{k=1}^n \sum_{j \le l} g^{(k)}_{jl} \alpha_k e^{-i(\Omega_{jl}^{(k)} - \omega_j - \omega_l)t - i\phi_k} \hat{a}_j^\dagger \hat{a}_l^\dagger + \mathrm{H.c.}
\end{equation}
\nomenclature{$\hat{H}_I(t)$}{Interaction Hamiltonian in the interaction picture respect to $\hat{H}_0$\nomrefeqpage}
\nomenclature{$\alpha_k$}{Amplitude of the $k$-th pump tone\nomrefeqpage}
\nomenclature{$\phi_k$}{Phase of the $k$-th pump tone\nomrefeqpage}
The resonance condition \( \Omega_{jl}^{(k)} \approx \omega_j + \omega_l \) selects the relevant processes contributing to pair creation in modes \( j \) and \( l \). Each tone drives a specific Bogoliubov transformation depending on its frequency and phase.

\subsection{Time Evolution and Bogoliubov Transformations}

We work in the interaction picture defined by the free Hamiltonian \( \hat{H}_0 \). The time evolution operator in this picture is
\begin{equation}
    \hat{U}_I(t) = \mathcal{T} \exp\left( -i \int_0^t \hat{H}_I(t') \mathrm{d} t' \right),
\end{equation}
\nomenclature{$\hat{U}_I(t)$}{Time evolution operator in the interaction picture\nomrefeqpage}
\nomenclature{$\mathcal{T}$}{Time-ordering operator\nomrefeqpage}
where \( \mathcal{T} \) denotes time-ordering. In the case of weak coupling or short times, we can approximate the evolution as Gaussian and neglect higher-order time-ordering effects.
 The effective transformation of mode operators is linear and described by a Bogoliubov transformation.

Let us define the vector of annihilation and creation operators $\hat{\bm{A}}$ and the vector of output operators after time evolution $\hat{\bm{A}}_\mathrm{out}$:
\begin{equation}
    \hat{\bm{A}} = \begin{pmatrix}
        \hat{a}_1, \cdots , \hat{a}_m,  \hat{a}^\dagger_1, \cdots  ,\hat{a}^\dagger_m
    \end{pmatrix}^\top
    , \quad
    \hat{\bm{A}}_\mathrm{out} = \hat{U}_I^\dagger(t) \, \hat{\bm{A}} \, \hat{U}_I(t).
\end{equation}
\nomenclature{$\hat{\bm{A}}$}{Vector of annihilation and creation operators\nomrefeqpage}
\nomenclature{$\hat{\bm{A}}_\mathrm{out}$}{Vector of output operators after time evolution\nomrefeqpage}
This time evolution implements a symplectic transformation:
\begin{equation}
    \hat{\bm{A}}_\mathrm{out} = \mathbf{S} \hat{\bm{A}},
    \label{eq:Aout}
\end{equation}
\nomenclature{$\mathbf{S}$}{Symplectic matrix describing the Bogoliubov transformation\nomrefeqpage}
where the symplectic matrix

\( \mathbf{S} \in \mathrm{Sp}(2m,\mathbb{R}) \) takes the form:
\begin{equation}\label{eq:S_U_V}
    \mathbf{S} = \begin{pmatrix}
    \mathbf{U} & \mathbf{V} \\
    \mathbf{V}^* & \mathbf{U}^*
    \end{pmatrix}, \quad \text{with } \mathbf{U} \mathbf{U}^\dagger - \mathbf{V} \mathbf{V}^\dagger = \mathbb{I},
\end{equation}
\nomenclature{$\mathbf{U}$}{Submatrix of $\mathbf{S}$ is the local squeezing matrix\nomrefeqpage}
\nomenclature{$\mathbf{V}$}{Submatrix of $\mathbf{S}$ is the inter-mode squeezing matrix\nomrefeqpage}

recovering the matrix form of the common input-output relation 
\begin{equation}
    \hat{a}_{\mathrm{out},i} = \sum_j\left(\mathbf{U}_{ij} a_i + \mathbf{V}_{ij} \hat{a}_j^\dagger\right).
\end{equation}

The Bogoliubov matrices $\mathbf{U}$ is the and $\mathbf{V}$ describe how a Gaussian unitary transformation mixes the input bosonic modes. $\mathbf{U}$ is the local squeezing matrix and
captures linear mode mixing, while 
$\mathbf{V}$ is the inter-mode squeezing matrix and captures squeezing and pair creation effects.

In the regime where each tone \( k \) is near-resonant with a pair \( (\omega_j + \omega_l) \), the corresponding contribution to the interaction Hamiltonian is approximately time-independent in the rotating frame. We can then write the total Hamiltonian as:
\begin{equation}
\label{eq:squeezparam}
    \hat{H}_\mathrm{eff} = \sum_{k=1}^n \sum_{j \le l} \left( r_{jl}^{(k)} \hat{a}_j^\dagger \hat{a}_l^\dagger + r_{jl}^{(k)*} \hat{a}_j \hat{a}_l \right),
\end{equation}
\nomenclature{$r_{jl}^{(k)}$}{Effective squeezing parameter from pump $k$ between modes $j$ and $l$\nomrefeqpage}
where we define the effective squeezing parameters as:
\begin{equation}
    r_{jl}^{(k)} = g^{(k)}_{jl} \alpha_k e^{-i\phi_k}.
\end{equation}

The total Hamiltonian is quadratic in the field operators and generates a Gaussian transformation.
Matrix $\mathbf{V}$ is directly related to the sum of all $r^{(k)}_{jl}$, allowing arbitrary interference between multiple pump tones with different phases.

The resulting symplectic transformation can be computed via matrix exponentiation:
\begin{equation}
    \mathbf{S}(t) = \exp\left( \bm{\Omega} \cdot \textbf{G} \cdot t \right),
\end{equation}
where 
\begin{equation}
\label{eq:symplectic}
    \bm{\Omega} = \begin{pmatrix}
    \mathbf{0}_{m} & \mathbb{I}_{m} \\
    -\mathbb{I}_{m} & \mathbf{0}_{m}
    \end{pmatrix}
\end{equation}
is the symplectic form and
\begin{equation}
    \textbf G \;=\;
\begin{pmatrix}
0 & \;\mathbf{G}\; \\
\mathbf{G}^* & 0
\end{pmatrix},
\quad
(\mathbf{G})_{jl} \;=\;\sum_{k=1}^n r_{jl}^{(k)}
\end{equation}
is the generator associated with \( \hat{H}_\mathrm{eff} \). 

$\mathbf{S}(t)$ maps the vector of input operators to that of the output operators, preserving all commutation relations. From $\mathbf{S}(t)$ one directly extracts the squeezing parameters and inter‑mode correlations, enabling both analytical and numerical evaluation of the output state’s squeezing and entanglement properties.

\nomenclature{\( \bm{\Omega} \)}{Symplectic form\nomrefeqpage}
\nomenclature{$\hat{G}$}{Matrix generator associated with the effective Hamiltonian\nomrefeqpage}

\subsection{Covariance Matrix and Output State}
\label{sect:Covariance Matrix and Output State}

To characterize the quantum state of the output field, we define the vector of quadrature operators in block-wise order:
\begin{equation}
    \hat{\bm{X}} = 
    \begin{pmatrix}
        \hat{x}_1, \cdots , \hat{x}_m,  \hat{p}_1, \cdots  ,\hat{p}_m
    \end{pmatrix}^\top
\end{equation}
\nomenclature{$\hat{\bm{X}}$}{Vector of quadrature operators \nomrefeqpage}
where
\begin{equation}
\label{eq:quadratures}
    \hat{x}_j = \frac{1}{\sqrt{2}} \left( \hat{a}_j + \hat{a}_j^\dagger \right), \quad
    \hat{p}_j = \frac{1}{\sqrt{2} i} \left( \hat{a}_j - \hat{a}_j^\dagger \right).
\end{equation}

\nomenclature{$\hat{x}_j$}{Position quadrature of mode $j$\nomrefeqpage}
\nomenclature{$\hat{p}_j$}{Momentum quadrature of mode $j$\nomrefeqpage}
These quadratures obey the block-wise commutation relations:
\begin{equation}\label{eq:symplecticOmega_block-wise}
    [\hat{X}_j, \hat{X}_k] = i \Omega_{jk}.
\end{equation}
where $\Omega_{jk}$ are the elements of the symplectic matrix defined in Eq. \ref{eq:symplectic}.

The state of the system is fully characterized by the first moments \( \langle \hat{\bm{X}} \rangle \) and the covariance matrix \( \bm{\sigma} \), defined by:
\begin{equation}
    \sigma_{jk} = \frac{1}{2} \langle \hat{X}_j \hat{X}_k + \hat{X}_k \hat{X}_j \rangle - \langle \hat{X}_j \rangle \langle \hat{X}_k \rangle.
\end{equation}

For a quadratic Hamiltonian and an initial vacuum state (a Gaussian state), the system evolves within the space of Gaussian states. The time-evolution operator \( \hat{U}_I(t) \) induces a symplectic transformation:
\begin{equation}
    \hat{\bm{X}}_\mathrm{out} = \mathbf{S}_X \hat{\bm{X}}, \qquad
    \bm{\sigma} = \mathbf{S}_X \bm{\sigma}_\mathrm{in} \mathbf{S}_X^\top,
\end{equation}
where \( \mathbf{S}_X \) is a real symplectic matrix acting on quadrature space and $\bm{\sigma}_\mathrm{in}$ is the input state covariance matrix.
\nomenclature{$\mathbf{S}_X$}{Symplectic matrix acting on quadrature space\nomrefeqpage}
\nomenclature{$\bm{\sigma}_\mathrm{in}$}{Input state covariance matrix\nomrefeqpage}
For the vacuum input:
\begin{equation}
\label{vac}
    \bm{\sigma}_\mathrm{in} =\bm{\sigma}_\mathrm{vac} = \frac{1}{2} \mathbb{I}_{2m}.
\end{equation}
\nomenclature{$\bm{\sigma}_\mathrm{vac}$}{Vacuum state covariance matrix\nomrefeqpage}
\nomenclature{$\mathbb{I}_{\textit{dim}}$}{Identity matrix of size \textit{dim}.}
Therefore, the output covariance matrix becomes:
\begin{equation}\label{eq:sigma_out_mode-wise}
    \bm{\sigma} = \frac{1}{2} \mathbf{S}_X \mathbf{S}_X^\top.
\end{equation}
\nomenclature{$\bm{\sigma}$}{Output covariance matrix\nomrefeqpage}
The quadrature operators are related to the mode operators via the linear transformation:
\begin{equation}
    \hat{\bm{X}} = \mathbf{T} \hat{\bm{A}}, \qquad
    \mathbf{T} = \frac{1}{\sqrt{2}}
    \begin{pmatrix}
    \mathbb{I}_m & -i \mathbb{I}_m \\
    \mathbb{I}_m & i \mathbb{I}_m
    
    \end{pmatrix}.
\end{equation}

Given the Bogoliubov transformation acting on mode operators, defined in Eq.~\eqref{eq:Aout} and Eq.~\eqref{eq:S_U_V}, the induced symplectic transformation in block-wise space is:
\begin{equation}
    \mathbf{S}_X = \mathbf{T} \, \mathbf{S} \, \mathbf{T}^{-1}.
\end{equation}

Carrying out this matrix product yields:
\begin{equation}
    \mathbf{S}_X = \begin{pmatrix}
    \mathrm{Re}(\mathbf{U} + \mathbf{V}) & -\mathrm{Im}(\mathbf{U} - \mathbf{V}) \\
    \mathrm{Im}(\mathbf{U} + \mathbf{V}) & \mathrm{Re}(\mathbf{U} - \mathbf{V})
    \end{pmatrix},
\end{equation}
where all blocks are \( m \times m \) real matrices.

Substituting into Eq.~\eqref{eq:sigma_out_mode-wise}, the final expression for the output covariance matrix is:
\begin{equation}
    \bm{\sigma} = \frac{1}{2}
    \begin{pmatrix}
    \mathrm{Re}(\mathbf{U} + \mathbf{V}) & -\mathrm{Im}(\mathbf{U} - \mathbf{V}) \\
    \mathrm{Im}(\mathbf{U} + \mathbf{V}) & \mathrm{Re}(\mathbf{U} - \mathbf{V})
    \end{pmatrix}
    \begin{pmatrix}
    \mathrm{Re}(\mathbf{U} + \mathbf{V})^\top & \mathrm{Im}(\mathbf{U} + \mathbf{V})^\top \\
    -\mathrm{Im}(\mathbf{U} - \mathbf{V})^\top & \mathrm{Re}(\mathbf{U} - \mathbf{V})^\top
    \end{pmatrix}.
\end{equation}

This real, symmetric \( 2m \times 2m \) matrix fully describes the Gaussian output state in mode-wise order and is directly compatible with the standard symplectic structure used in continuous-variable quantum information protocols.

\subsubsection*{Assumption 1: Uniform Single-Mode Squeezing.}

We consider a configuration in which each mode of a multimode transmission line undergoes identical, uncorrelated single-mode squeezing. This can be approximately realized using a multi-pump traveling-wave parametric amplifier (TWPA), where each mode is driven by a separate pump tone tuned to twice its frequency (degenerate parametric amplification). To ensure uniform squeezing across all modes, the following conditions must be met:
\begin{itemize}
  \item a flat gain profile across the operational bandwidth,
  \item negligible dispersion along the transmission line,
  \item stable phase coherence among the pump tones.
\end{itemize}
These engineering requirements can be satisfied via spectral pump shaping and precise calibration of the transmission line properties.

Under these conditions, all modes experience equal squeezing strength, defined by the modulus of the squeezing coefficient \( |r| \) in Eq. \ref{eq:squeezparam}, and aligned squeezing angles. In this regime, the Bogoliubov matrices take a simple diagonal form:
\begin{equation}
\label{eq:UVfromr}
\mathbf{U} = \cosh |r| \cdot \mathbb{I}_m, \qquad \mathbf{V} = \sinh|r|\cdot \mathbb{I}_m,
\end{equation}

Each \( (x_j, p_j) \) pair is independently squeezed. The variances are:
\begin{equation}
\mathrm{Var}(x_j) = \frac{1}{2} e^{-2|r|}, \qquad \mathrm{Var}(p_j) = \frac{1}{2} e^{2|r|}.
\end{equation}

\subsubsection*{Assumption 2: Pure Two-Mode Squeezing Between Arbitrary Modes.}

We now consider ideal two-mode squeezing between an arbitrary pair of modes \( j \) and \( l \) (with \( j \neq l \)), while all other modes remain in vacuum. This can be realized in a multi-pumped TWPA architecture by applying a pump tone at frequency \( \Omega_{jl} = \omega_j + \omega_l \), enabling non-degenerate parametric down-conversion via three-wave mixing. If coupling to other modes is negligible—achieved through spectral sparsity, dispersion engineering, and pump frequency selection—the effective interaction is confined to the selected pair \( (j,l) \).

To realize clean, isolated two-mode squeezing in practice, several physical conditions must be satisfied:
\begin{itemize}
  \item \textbf{Spectral selectivity:} The mode spectrum should be sufficiently sparse or engineered to avoid accidental degeneracies, ensuring that each pump addresses only its intended mode pair.
  \item \textbf{Minimal dispersion:} The transmission line should exhibit low dispersion across the operating bandwidth to maintain consistent phase matching across the selected frequency pairs.
  \item \textbf{Pump coherence and stability:} The pump tones must be phase-stable and frequency-precise to prevent drift or cross-coupling between unintended modes.
\end{itemize}
These requirements can be addressed through careful design of the TWPA geometry, spectral shaping of the pump distribution, and precise calibration of system parameters. When these conditions are met, the squeezing interaction is effectively localized to the selected mode pair, and the squeezing strengh $|r|$ can be assumed uniform and the squeezing angles aligned across all pairs.

Under these assumptions, the Bogoliubov matrices \( \mathbf{U} \) and \( \mathbf{V} \) acquire a block structure determined by the mode connectivity. To capture this structure explicitly, we define the symmetric binary adjacency matrix \( \mathbf{A} \in \{0,1\}^{m \times m} \). 
\nomenclature{$\mathbf{A}$}{Binary adjacency matrix\nomrefpage}
For a pure two-mode squeezed state coupling only modes \( j \) and \( l \), the matrix \( \mathbf{A} \) is zero everywhere except:
\[
\mathbf{A}_{jl} = \mathbf{A}_{lj} = 1.
\]
The Bogoliubov matrices become:
\begin{equation}
\label{eq:uandv}
    \mathbf{U} = \mathbb{I}_m + (\cosh |r| - 1)\cdot \mathrm{diag}(\mathbf{D}), \qquad \mathbf{V} = \sinh |r| \cdot \mathbf{A},
\end{equation}

where \( \mathbf{D} \) is a diagonal matrix encoding mode participation in two-mode squeezing processes. Specifically, \( \mathbf{D}_{jj} = 1 \) if mode \( j \) is coupled to at least one other mode via a nonzero entry in \( \mathbf{A} \), and \( \mathbf{D}_{jj} = 0 \) otherwise. That is,
\[
\mathbf{D}_{jj} = 
\begin{cases}
1, & \text{if } \sum_{l=1}^{m} \mathbf{A}_{jl} > 0, \\
0, & \text{otherwise}.
\end{cases}
\]

This construction ensures that each mode involved in squeezing undergoes the correct Bogoliubov transformation gain in \( \mathbf{U} \), while modes that are not coupled remain unaffected.

In the simple case of a single pair \( (j,l) \), the matrix \( \mathbf{A} \) has value 1 at \( (j,l) \) and \( (l,j) \), and zeros elsewhere. This formulation generalizes naturally to arbitrary graphs of squeezed pairs: any symmetric adjacency matrix \( \mathbf{A} \) with entries in \( \{0,1\} \) encodes which modes are pairwise entangled via two-mode squeezing. The resulting matrices \( \mathbf{U} \) and \( \mathbf{V} \) then describe a multimode Gaussian state with an entanglement structure.

These matrices are real and symmetric, implying:
\[
\mathrm{Im}(\mathbf{U} \pm \mathbf{V}) = 0, \qquad
\mathrm{Re}(\mathbf{U} \pm \mathbf{V}) = \mathbf{U} \pm \mathbf{V}.
\]

Constructing the real symplectic transformation \( \mathbf{S}_X \) in mode-wise quadrature order:
\begin{equation}
\mathbf{S}_X = \begin{pmatrix}
\mathbf{U} + \mathbf{V} & 0 \\
0 & \mathbf{U} - \mathbf{V}
\end{pmatrix},
\end{equation}
we find the output covariance matrix as:
\begin{equation}
\label{eq:sigmaoutsimb}
\boldsymbol{\sigma} = \frac{1}{2} \, \mathbf{S}_X \, \mathbf{S}_X^\top = \frac{1}{2}
\begin{pmatrix}
(\mathbf{U} + \mathbf{V})^2 & 0 \\
0 & (\mathbf{U} - \mathbf{V})^2
\end{pmatrix}.
\end{equation}

\subsection{Cluster States and Nullifiers}

Cluster states are a central resource for continuous-variable (CV) measurement-based quantum computation. These multipartite entangled Gaussian states are defined, in the infinite squeezing limit, as simultaneous zero-eigenstates of a set of commuting observables called \emph{nullifiers}. They generalize the stabilizer formalism of discrete-variable cluster states to the CV setting.

\subsubsection{Nullifiers and Adjacency Matrix Representation}

A CV cluster state associated with a symmetric adjacency matrix \( \mathbf{A} \) is ideally characterized by the vanishing of the nullifier operators.
For a given cluster state defined by an adjacency matrix, the ideal nullifier operator associated with node \( j \) takes the form:
\begin{equation}\label{eq:null_j}
    \hat{\delta}_j = \hat{p}_j - \sum_{l=1}^m A_{jl} \, \hat{x}_l, \qquad j = 1, \dots, m.
\end{equation}
where \( \hat{x}_l \) and \( \hat{p}_j \) are the position and momentum quadratures of the modes.
The nullifier operator couples the momentum quadrature of a given mode to the position quadratures of its graph neighbors, thereby reflecting the entanglement structure imposed by the graph topology.

In the block-wise quadrature basis, the nullifier vector takes the form:
\begin{equation}
    \hat{\bm{\delta}} = \mathbf{N} \cdot \hat{\bm{X}}, \qquad \text{with} \quad \mathbf{N} = \left( -\mathbf{A} \;\middle|\; \mathbb{I}_m \right).
\end{equation}
\nomenclature{$\hat{\bm{\delta}}$}{Vector of the nullifier operators\nomrefeqpage}
\nomenclature{$\mathbf{N}$}{Nullifier matrix acting on the quadrature vector\nomrefeqpage}
This definition naturally encodes the entanglement structure of the target cluster state.
In the ideal limit of infinite squeezing, the variances of the nullifiers vanish, and the state becomes fully supported on the common eigenspace of the nullifiers.

\subsubsection{Covariance-Based Characterization and Connection to the Bogoliubov Formalism}

Let \( \bm{\sigma} \in \mathbb{R}^{2m \times 2m} \) denote the covariance matrix of the output Gaussian state. The noise in the nullifier observables is quantified by the nullifier covariance matrix:
\begin{equation}
    \bm{\Delta} = \mathbf{N} \bm{\sigma} \mathbf{N}^\top,
\end{equation}
\nomenclature{$\bm{\Delta}$}{Nullifier covariance matrix\nomrefeqpage}
where each diagonal entry \( \Delta_{jj} \) corresponds to the variance of the nullifier operator \( \hat{\delta}_j \), i.e.,
\[
\Delta_{jj} = \langle \hat{\delta}_j^2 \rangle - \langle \hat{\delta}_j \rangle^2 = \mathrm{Var}(\hat{\delta}_j).
\]
This directly quantifies how closely the physical state approximates the ideal cluster condition \( \hat{\delta}_j = 0 \), and allows a mode-by-mode assessment of the squeezing and entanglement structure. In particular, the eigenvalues of \( \bm{\Delta} \) quantify the noise present in orthogonal linear combinations of the nullifier operators. Smaller eigenvalues correspond to directions in nullifier space where the noise (i.e., variance) is minimized, indicating that the corresponding combinations are closer to ideal quantum correlations. Since \( \bm{\Delta} \) is a real, symmetric, and positive semi-definite matrix, its eigenvalues can be interpreted as generalized variances along principal axes defined by a suitable basis transformation. This diagonalization effectively identifies the most and least squeezed (i.e., quantum) collective modes of the system. When all eigenvalues are significantly below the vacuum noise level (e.g., \( 1/2 \) in units where \( \hbar = 1 \)), the state exhibits genuine multipartite entanglement and closely approximates the ideal cluster state. Therefore, analyzing the spectrum of \( \bm{\Delta} \) provides a basis-independent and physically meaningful method to assess the quantumness and fidelity of the generated state.

\subsubsection*{Example: Tripartite Cyclic Cluster State (Ideal Squeezing)}

As a concrete example, consider a tripartite cluster state with a cyclic topology, where each of the three modes is connected to the other two. The corresponding binary adjacency matrix \( \mathbf{A} \in \{0,1\}^{3 \times 3} \) is:

\begin{equation}
    \mathbf{A} = 
    \begin{pmatrix}
        0 & 1 & 1 \\
        1 & 0 & 1 \\
        1 & 1 & 0
    \end{pmatrix}.
\end{equation}

The ideal nullifiers for this state take the form:
\begin{align}
    \hat{\delta}_1 &= \hat{p}_1 - \hat{x}_2 - \hat{x}_3, \\
    \hat{\delta}_2 &= \hat{p}_2 - \hat{x}_1 - \hat{x}_3, \\
    \hat{\delta}_3 &= \hat{p}_3 - \hat{x}_1 - \hat{x}_2.
\end{align}

In the infinite squeezing limit, an ideal cluster state satisfies
\begin{equation}
    \hat{\delta}_j |\psi_\mathrm{cluster}\rangle = 0 \qquad \text{for } j = 1,2,3.
\end{equation}

This means the quantum state \( |\psi_\mathrm{cluster}\rangle \) lies in the simultaneous zero-eigenspace of the three nullifiers. The vanishing of their variances,
\begin{equation}
    \langle \hat{\delta}_j^2 \rangle \rightarrow 0 \quad \text{as} \quad |r| \rightarrow \infty,
\end{equation}
indicates perfect entanglement and full compliance with the cluster-state stabilizer conditions. In realistic scenarios with finite squeezing, these variances remain small but nonzero, and their values serve as fidelity indicators with respect to the ideal state.


In the case of pure two-mode squeezing, the matrices \( \mathbf{U} \) and \( \mathbf{V} \) were shown to be real and symmetric, with their structure determined by matrix \( \mathbf{A} \), indicating physical connectivity:
\begin{equation}
    \mathbf{V} = \sinh |r| \cdot \mathbf{A}.
\end{equation}

While \( \mathbf{V} \) describes which mode pairs are entangled through squeezing operations, \( \mathbf{A} \) defines the target cluster state graph in block-wise space. These matrices are related conceptually, but not necessarily equal: \( \mathbf{V} \) encodes the actual implementation of squeezing interactions, while \( \mathbf{A} \) describes the structure of nullifiers and desired stabilizers.

In optimized designs, one seeks to engineer \( \mathbf{V} \) such that the resulting \( \bm{\sigma} \) leads to nullifier variances that are minimized for a chosen \( \mathbf{A} \). This can be accomplished by tuning the pump tones to induce a pattern of effective squeezing coefficients \( r_{jl}^{(k)} \) that sum (via Eq.~\eqref{eq:S_U_V}) into a \( \mathbf{V} \) approximately proportional to \( \mathbf{A} \), or via more general linear optical transformations.

\subsubsection{Performance Metrics}

To assess the quality of an approximate cluster state, one can compute the following figures of merit from the nullifier covariance matrix \( \bm{\Delta} \), normalized with respect to the structure-dependent shot noise limit (SNL). This normalization enables fair comparison across different graph topologies defined by the adjacency matrix \( \mathbf{A} \).

From the nullifier operators definition in Eq.~\eqref{eq:null_j} and assuming that all modes are initialized in vacuum (i.e., \( \langle \hat{x}_l^2 \rangle = \langle \hat{p}_j^2 \rangle = 1/2 \), and all covariances vanish), the variance of the nullifier is:
\begin{equation}
    \langle \hat{\delta}_j^2 \rangle = \langle \hat{p}_j^2 \rangle + \sum_l A_{jl}^2 \langle \hat{x}_l^2 \rangle = \frac{1}{2} \left( 1 + \sum_l A_{jl}^2 \right).
\end{equation}

This expression captures how vacuum fluctuations propagate through the cluster state’s graph structure. A similar derivation appears, for example, in the theoretical framework described by Menicucci et al.~\cite{menicucci2006universal} and further discussed in Gaussian cluster-state literature (e.g., \cite{Alexander2018}).

Using this, we define the

\begin{itemize}
    \item \textbf{normalized average nullifier variance}:
    \begin{equation}\label{Normalized average nullifier variance}
        \langle \hat{\delta}^2 \rangle_{\mathrm{avg}}^{\text{(norm)}} = \frac{1}{m} \sum_{j=1}^m \frac{ 2\bm{\Delta}_{jj}}{1 + \sum_l A_{jl}^2},
    \end{equation}
\nomenclature{$\langle \hat{\delta}^2 \rangle_{\mathrm{avg}}^{\text{(norm)}}$}{Normalized average nullifier variance\nomrefeqpage}

\end{itemize}

which accounts for the vacuum noise scaling imposed by the cluster state's graph connectivity. The normalized variances directly indicate how much squeezing is achieved relative to the ideal limit, and values below 1 imply entanglement in the corresponding nullifier mode. Meanwhile, the Frobenius distance provides a global, quantitative measure of state closeness, incorporating all quadrature correlations simultaneously.

This nullifier-based formalism thus bridges experimental observables (via \( \mathbf{V} \) and \( \bm{\sigma} \)) with the graph-theoretic structure of ideal cluster states (encoded in \( \mathbf{A} \)). In the context of multi-pumped TWPAs, these normalized metrics support the systematic verification of programmable cluster topologies and provide a robust benchmark for scalable measurement-based quantum information processing.\\

Another possible performance metric is the

\begin{itemize}

   \item \textbf{Frobenius distance:}
   
   \begin{equation}\label{eq:frobenius}
        \widetilde{{\mathcal{{F}}}}_\text{{F}}(\sigma, \Gamma) = 
        \exp\left( -\frac{ \left\| \sigma - \Gamma \right\|_\text{{F}}^2 }{ \left\| \sigma \right\|_\text{{F}}^2 } \right),
    \end{equation}
\nomenclature{$\widetilde{{\mathcal{{F}}}}_\text{{F}}$}{Frobenius distance\nomrefeqpage}
   where \( \bm{\Gamma} \) and \( \bm{\sigma} \) are both \( 2m \times 2m \) covariance matrices, which measures the similarity between the experimentally generated Gaussian state (\( \bm{\Gamma} \)) and the ideal target cluster state (\( \bm{\sigma} \)). The ideal covariance matrix \( \bm{\sigma} \) can be constructed by applying the appropriate symplectic transformation associated with the adjacency matrix \( \mathbf{A} \) to the vacuum state.

\end{itemize}

\newpage

\section{Reference State Protocol and Covariance Matrix Reconstruction}

In typical continuous-variable experiments, the measurement of quadratures via homodyne or heterodyne detection requires a well-defined phase reference, set by the LO. Similarly, the generation of squeezed states via parametric processes depends on the phase of the pump field, which defines the squeezing angle. In the absence of a stable or known phase relation between the pump and the LO, the measured quadratures are rotated by unknown phases \( \theta_i \), and the observed statistics correspond to an ensemble over these unknown local phases.

However, it is still possible to reconstruct the covariance matrix of the output Gaussian state $\bm{\Gamma}$\nomenclature{$\bm{\Gamma}$}{Reconstructed output covariance matrix\nomrefeqpage} by exploiting the structure of quantum correlations. This approach provides a data-driven method to self-calibrate the phase reference.

\subsection{Scaled Covariance Matrix}
In the notation defined in Eq. \ref{eq:quadratures}, the raw voltage data of the $j$-th mode ($I_j$, $Q_j$) acquired from the measurement system are scaled to quadrature operators expressed in photon-units ($i_j$, $q_j$) via the relations:
\nomenclature{$I_{j}$}{$j$-th mode raw voltage position quadrature\nomrefeqpage}
\nomenclature{$Q_{j}$}{$j$-th mode raw voltage momentum quadrature\nomrefeqpage}

\begin{equation}
i_{j}^{}= \frac{I_j}{\sqrt{2G_j Z_0 \hbar\omega_j/2 \tau^{-1}}}
, \quad 
q_{j}^{}= \frac{Q_j}{\sqrt{2G_j Z_0 \hbar\omega_j \tau^{-1}}}
\end{equation}
\nomenclature{$i_{j}$}{$j$-th mode measured position quadrature in photon-number units\nomrefeqpage}
\nomenclature{$q_{j}$}{$j$-th mode measured momentum quadrature in photon-number units\nomrefeqpage}
\nomenclature{$G_{j}$}{$j$-th mode system gain\nomrefeqpage}
\nomenclature{$Z_{0}$}{Characteristic impedance of the transmission line\nomrefeqpage}
\nomenclature{$\tau$}{Acquisition time\nomrefeqpage}

where $\theta_j$ is the unknown rotation angle of the $j$-th spectral mode, $G_j$ denotes the system gain for the $j$-th spectral mode, $Z_0 = 50\ \Omega$ is the characteristic impedance of the transmission line, $\omega_j/2\pi$ is the mode frequency, and $\tau = 6\text{ }\mu\text{s}$ is the acquisition time.

From repeated measurements, we construct the measured scaled covariance matrix elements:

\begin{equation}
    \Gamma^{\text{exp}}_{j,l} = \frac{1}{2} \left \langle {M}_{j} {M}_{l} + {M}_{l} {M}_{j} \right \rangle - \left \langle {M}_{j} \right \rangle \left \langle {M}_{l} \right \rangle.
\end{equation}
\nomenclature{$\bm{\Gamma}^{\mathrm{exp}}$}{Scaled measured covariance matrix\nomrefeqpage}
where \( {\bm{M}}_{\bm{}}=({i}_1^{}, {q}_1^{}, \dots, {i}_m^{}, {q}_m^{})^T \).
\nomenclature{$\bm{M}$}{Measured quadratures vector in photon basis\nomrefeqpage}
This construction of the $M$ vector is in a mode-wise notation.
The covariance matrix in the yet unknown $\theta$-rotated basis $\bm{{\Gamma}_{\theta}}$ of the two-mode quantum state generated at the output of the TWPA is reconstructed following a reference state subtraction method \cite{PhysRevLett.107.113601}:

\begin{equation}
\label{exp:gamma}
    \bm{\Gamma_{\theta}} = (\bm{\Gamma^{\text{exp}}_{\mathrm{ON}}} -\bm{\Gamma^{\text{exp}}_{\mathrm{OFF}}}) + \bm{{\sigma^{\text{in}}}},
\end{equation}
\nomenclature{$\bm{\Gamma}_{\theta}$}{Rotated reconstructed output covariance matrix\nomrefeqpage}

where $\bm{\Gamma^{\text{exp}}_{\mathrm{ON}}}$ and $\bm{\Gamma^{\text{exp}}_{\mathrm{OFF}}}$ are the scaled covariance matrices measured with the pumps on and off, respectively, and $\bm{\sigma}_\mathrm{in} = \frac{1}{2} \bm{\mathbb{I}}\coth{\frac{\hbar \omega_j}{2k_bT_j}}$ is the covariance matrix of the input thermal state, which reduces to the vacuum state defined in Eq. \ref{vac} in the limit for $T_{j}\rightarrow0$.
Since the choice on the LOs phases is arbitrary, the reconstructed covariance matrix $\bm{\Gamma_{\theta}}$ is in a yet undefined $\theta$-rotated basis $\bm{{X}}_{\bm{\theta}}^{(\text{exp})}$ such that

\begin{equation}\label{eq:rotmat}
    {\bm{X}}_{\bm{\theta}}^{(\text{exp})} = \mathbf{R}_{\bm{\theta}} \cdot {\bm{X}}^{(\text{exp})}, 
\qquad
\bm{\Gamma_\theta = \textbf{R}_{\theta} \cdot \Gamma  \cdot\textbf{R}_{\theta}^{\text{T}}},
\qquad
    \bm{R}_{\bm{\theta}} =\bigoplus_{j=1}^m 
    \begin{pmatrix}
    \cos{\theta_j} & -\sin{\theta_j}\\
    \sin{\theta_j} & \cos{\theta_j}
    \end{pmatrix}.
\end{equation}
\nomenclature{$\textbf{R}_{\theta}$}{Rotation matrix\nomrefeqpage}
\nomenclature{$\bm{X}_{\bm{\theta}}$
}{Rotated vector of quadrature operators\nomrefeqpage}
where ${\bm{X}}_{\bm{}}^{\text{exp}}$ is the experimental cluster state basis expressed in the mode-wise notation.
However, by numerically inverting Eq. \ref{eq:rotmat}, we can reconstruct the output covariance matrix.

\subsection{System Gain Calibration}

The system gain $G_j$ and system noise temperature $T_\mathrm{SYS,j}$ for each spectral mode are calibrated by measuring the noise power spectral density at the output of a variable temperature stage (VTS) as a function of its temperature $T_\mathrm{VTS}$.

The measured noise power is fitted to the following expression where $G_j$ and $T_\mathrm{SYS,j}$ are the unknown parameters:

\begin{equation}
\frac{\langle I_j^2 + Q_j^2 \rangle}{Z_0 \tau_j^{-1}} = G_j \left[ \frac{\hbar \omega_j}{2} \coth\left(\frac{\hbar \omega_j}{2 k_\text{B} T_\mathrm{VTS}}\right) + k_\text{B} T_\mathrm{SYS, j} \right].
\end{equation}
\nomenclature{$T_{\text{SYS},j}$}{$j$-th mode system noise temperature \nomrefeqpage}
\nomenclature{$T_{\text{VTS}}$}{Variable Temperature Stage (VTS) temperature\nomrefeqpage}

Here $I_j$ and $Q_j$ are the in-phase and quadrature voltage components, respectively, $\omega_j$ is the mode frequency, and $k_\text{B}$ is the Boltzmann constant.

\subsection{Gaussianity Validation}

To validate that the measured quantum states retain Gaussian character—a critical assumption underpinning cluster-state quantum computation with continuous variables—we performed a comprehensive statistical analysis on a representative subset of quadrature data collected via heterodyne detection. The analysis is here reported focusing, as an example, on the linear four-mode cluster configuration presented in Figure 3(a,e) of the main text and includes:

\begin{itemize}
    \item \textbf{Inspection of marginal histograms} to verify Gaussian-shaped distributions (see Figure~\ref{fig:HistogramQ}).
    \item \textbf{Moment analysis up to fourth order}, including skewness, excess kurtosis, and fourth-order cumulants (see Figure~\ref{fig:SkewnessEtc}).
    \item \textbf{Standard normality tests}, namely the Shapiro–Wilk and Anderson–Darling tests (see Figure~\ref{fig:Normality}).
\end{itemize}

\begin{figure}[h]
    \centering
    \includegraphics[width=\linewidth]{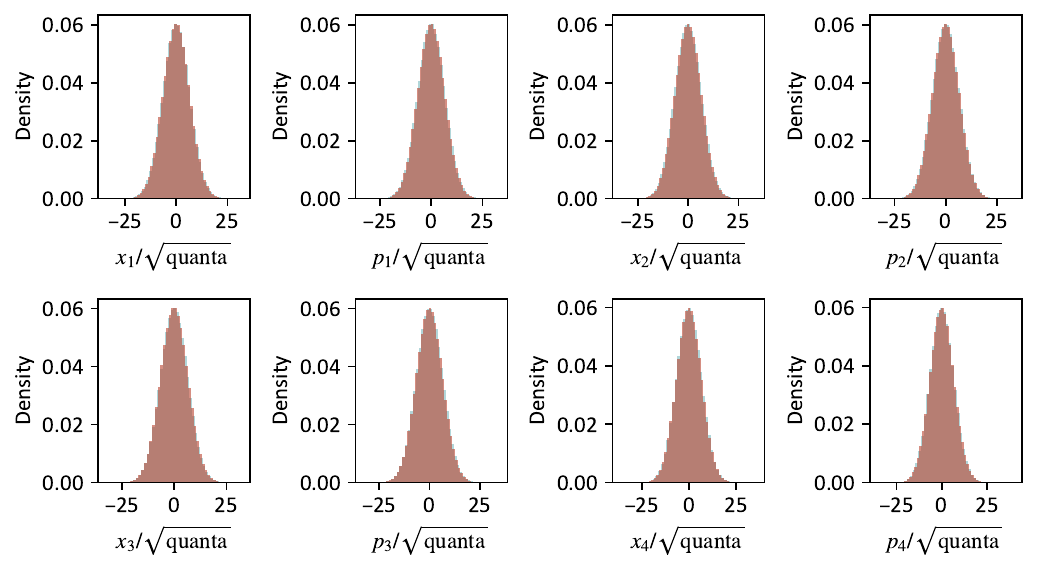}
    \caption{\textbf{Gaussian quadrature distributions of the four measured modes in the linear cluster state.}  
    Heterodyne detection of the four modes yields Gaussian distributions for each quadrature, consistent with the expected statistics of squeezed states forming the linear cluster. The data correspond to the same cluster configuration shown in Figure 3(a,e) of the main text. The measured variances are used to reconstruct the covariance matrix and assess entanglement through the nullifier variances.}
    \label{fig:HistogramQ}
\end{figure}

\begin{figure}[h]
    \centering
    \includegraphics[width=\linewidth]{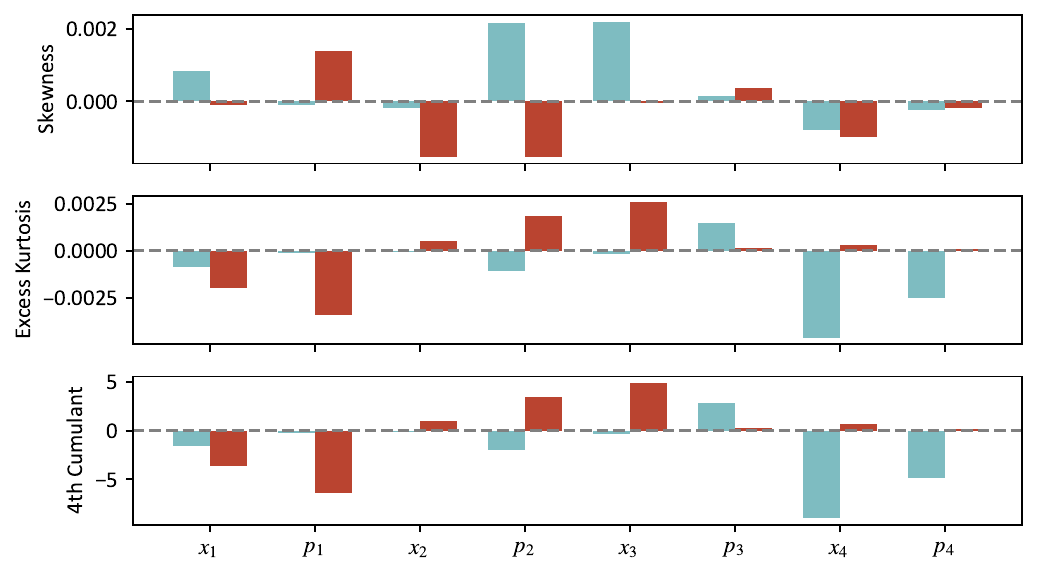}
    \caption{\textbf{Higher-order statistical moments of the measured quadratures for the linear cluster state.}  
    We report the skewness, excess kurtosis, and fourth-order cumulant of the quadrature distributions for the four measured modes, comparing the vacuum reference state (pump off, light blue) and the cluster state (pump on, red), as shown in Figure 3(a,e) of the main text. In all cases, the skewness remains below 0.002 and the excess kurtosis below 0.0025, confirming the near-Gaussian character of the distributions and validating the Gaussian cluster state assumption underlying the nullifier-based analysis.}
    \label{fig:SkewnessEtc}
\end{figure}

\begin{figure}[h]
    \centering
    \includegraphics[width=\linewidth]{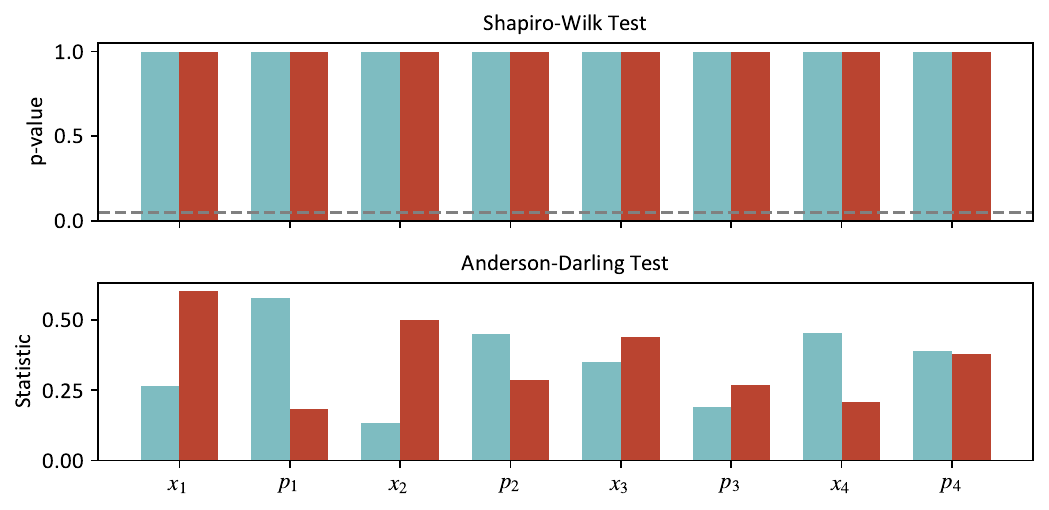}
    \caption{\textbf{Normality tests on quadrature distributions for the linear cluster state.}  
    We show the results of two standard normality tests—Shapiro–Wilk and Anderson–Darling—applied to the quadrature distributions of the four measured modes, for both the vacuum reference state (pump off, light blue) and the cluster state (pump on, red), corresponding to the configuration analyzed in Figure 3 (a,e) in the main text. The Shapiro–Wilk test yields p-values close to 1 for all modes, indicating no statistically significant deviation from normality. The Anderson–Darling test statistics remain between 0.25 and 0.5 in all cases, further supporting the Gaussian character of the measured states and justifying the use of covariance-based criteria in the entanglement analysis.}
    \label{fig:Normality}
\end{figure}

All results confirm that the measured quadrature distributions are statistically indistinguishable from ideal Gaussians within experimental uncertainty. This supports the validity of Gaussian-state formalism and covariance-based methods used throughout the entanglement and nullifier analysis.

\subsection{Multiplexed Heterodyne Detection}

The calibration and validation procedures were performed using a broadband heterodyne detection system with simultaneous acquisition of multiple frequency modes. LOs were carefully placed to avoid image band contamination and were phase-locked to ensure consistent quadrature references across the entire spectral span.

This multiplexed detection strategy is crucial to resolve all entangled modes in a single shot, enabling full reconstruction of the multimode covariance matrix with minimal overhead.

\subsection{Optimization Strategy: Maximizing Inter-mode Covariances}

Our optimization strategy consists in the maximization of $p_j-x_l$ correlations over the set of unknown local oscillator phases \( \bm{\theta} \), effectively determining the optimal global quadrature basis in which the reconstructed covariance matrix \( \bm{\Gamma}_{\theta} \) most closely resembles an ideal cluster state.
To this end, we maximize
\begin{equation}\label{eq:Cxp_definition}
    \mathcal{C}_{px}(\theta) = \frac{1}{m^2} \sum_{j=1}^{m} \sum_{l=1}^{m} \bm{\Gamma}_{\theta, p_j, x_l}
\end{equation}
\nomenclature{$\mathcal{C}_{px}(\theta)$}{Mean value x-p correlations\nomrefeqpage}
which is the mean value of all pairwise covariances between the \( p_j \) and \( x_l \) quadratures in the rotated covariance matrix \( \bm{\Gamma}_\theta \).

Crucially, the set of phases \( \bm{\theta} \) of the LOs obtained by this optimization are not arbitrary: they are directly related to the unknown pump phases used to generate squeezing in the source. These pump phases determine the orientation of the squeezing ellipse in phase space, i.e., the squeezing angle, and are typically not accessible through direct measurement. As a result, the effective squeezing observed via heterodyne detection depends on the alignment between the LO phase and the unknown squeezing angle. Optimizing \( \bm{\theta} \) thus serves not only to reveal the strongest squeezing present in the system, but also to compensate for unknown phase shifts introduced during state preparation.

\subsection{Reconstruction of the Covariance Matrix}
\label{subsec:optimize}

Once the optimal angle
\[
\bm{\theta}_\mathrm{opt} = \arg\max_{\theta} \mathcal{C}_{px}(\theta)
\]
\nomenclature{$\bm{\theta}_\mathrm{opt}$}{Vector of angles which maximize $\mathcal{C}_{px}(\theta)$ \nomrefeqpage}
is found, the covariance matrix in the original basis can be estimated by applying the inverse of the rotation matrix:
\begin{equation}
    \bm{\Gamma} = \mathbf{R}_{\bm{\theta}_\mathrm{opt}}^{-1} \bm{\Gamma}_{\theta} \left(\mathbf{R}_{\bm{\theta}_\mathrm{opt}}^{-1} \right)^\top,
    \label{eq:sigma_corrected}
\end{equation}

This procedure, for which we report the effects in Figure \ref{fig:7s}, effectively recovers a global phase reference, aligning both the measured quadratures and the squeezing direction. It is valid provided that the output state exhibits sufficient two-mode correlations.

\begin{figure}[h]
    \centering
    \includegraphics[width=\linewidth]{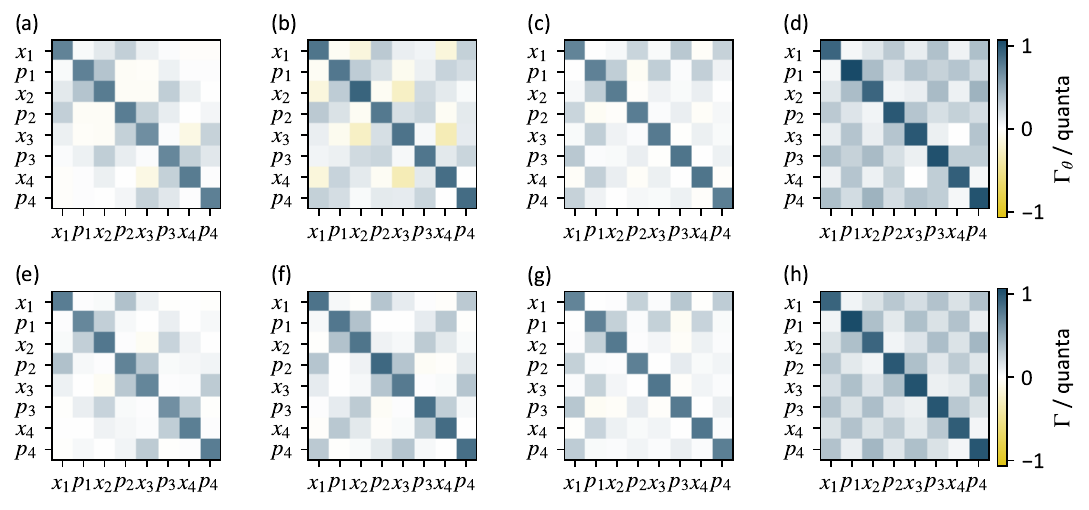}
    \caption{\textbf{Effect of the optimization strategy based on $p-x$ correlation maximization of $\bm{\Gamma_{\theta}}$.}  
    Covariance matrices reported in Figure 3 of the paper prior (a-d) and post (e-h) the optimization algorithm reported in Subsection \ref{subsec:optimize}.}
    \label{fig:7s}
\end{figure}

We then compared the reconstructed covariance matrix $\Gamma$ with its theoretical counterpart $\sigma$ using the Frobenius distance introduced in Eq. \ref{eq:frobenius}. The comparison is shown in Figure \ref{fig:8s}.

\begin{figure}[h]
    \centering
    \includegraphics[width=\linewidth]{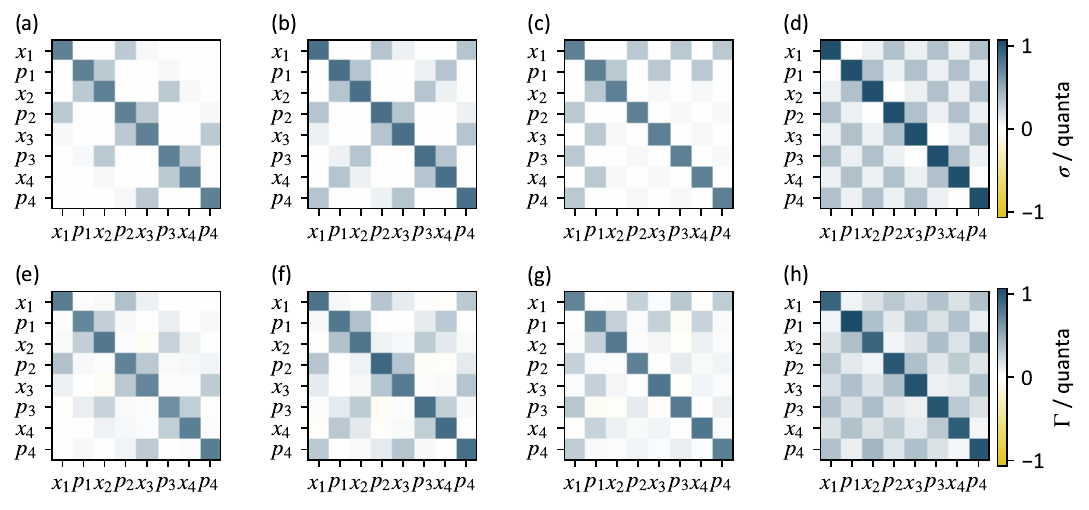}
    \caption{\textbf{Comparison between simulated and experimental covariance matrices.} Theoretical covariance matrices (panels a–d) correspond to simulations (see the Appendix) performed using the squeezing parameter \( r \) that maximizes the Frobenius fidelity defined in Eq.~\ref{eq:frobenius}. Experimental covariance matrices reconstructed from data are shown in panels (e–h). The optimized squeezing parameters used in the simulations are, respectively, \( |r| = 0.277, 0.283, 0.281, \) and \( 0.286 \) for panels (a) through (d). The maximum relative deviation of $|r|$ is below \(1.7\%\), confirming the robustness and consistency of the parameter estimation across different experiments.
}

    \label{fig:8s}
\end{figure}

\subsection{Discussion}

The reconstruction protocol presented here addresses the challenge of unknown quadrature rotations arising from phase delays accumulated during signal propagation through the readout chain. While the local oscillators and the pump tones are phase-locked to a common reference—ensuring overall coherence—the detection phases at different frequencies can experience distinct, frequency-dependent shifts due to dispersive and thermal effects in the measurement path. These shifts are typically unknown and difficult to calibrate directly.

By exploiting inter-mode quantum correlations, in particular those induced by two-mode squeezing, we implement a data-driven optimization strategy that compensates for these residual phase offsets. The optimal set of local quadrature angles is determined numerically by maximizing the off-diagonal correlations in the rotated covariance matrix \( \bm{\Gamma}_{\theta} \). This allows us to reconstruct the physical covariance matrix \( \bm{\Gamma} \) in a consistent global quadrature basis.

The validity of this reconstruction is confirmed by several independent checks: the observed quadrature statistics remain indistinguishable from ideal Gaussian distributions, and the reconstructed covariance matrices match closely with theoretical predictions. This ensures that key properties of the state, such as entanglement structure and nullifier variances, can be reliably extracted.

Importantly, this protocol requires no additional calibration hardware and is compatible with broadband, multiplexed detection schemes. It is therefore particularly suited to integrated or cryogenic platforms, where frequency-dependent phase delays are common and in-situ phase calibration is challenging. The method provides a robust and scalable approach to covariance matrix reconstruction for high-dimensional Gaussian states in realistic experimental settings.

\newpage

\appendix
\section{Examples of Covariance Matrices Analytical Calculations}

In practical implementations of multimode Gaussian cluster states using multi-pump traveling wave parametric amplifiers (TWPAs), it is common for \emph{ancillary modes}—modes not explicitly part of the target cluster graph—to become entangled with the main cluster modes via two-mode squeezing (TMS) processes. These ancillary couplings can significantly affect the output covariance matrix by modifying modal gain distributions and introducing residual correlations (see Figure \ref{fig:ancillary}).

\begin{figure}[h]
    \centering
    \includegraphics[width=0.5\linewidth]{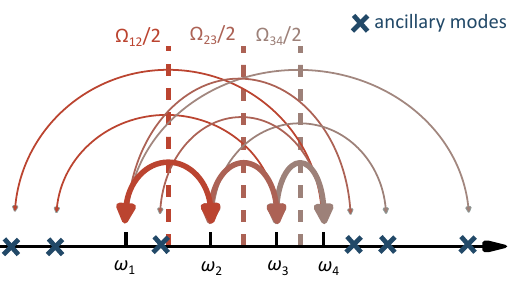}
    \caption{\textbf{Ancillary Modes in a Linear Cluster State.} Example of ancillary modes in a linear cluster state. Although these modes are not part of the intended cluster topology, they do participate in the amplification process due to their coupling with connected modes via the pump tones.}
    \label{fig:ancillary}
\end{figure}

For accurate fidelity estimation and nullifier optimization, it is therefore essential to include these ancillary modes in the theoretical modeling.

Coherently with the proposed theoretical framework, all simulations were carried out in block-wise notation and subsequently converted into the mode-wise notation by applying the transformation
\begin{equation}
    \bm{\sigma}_\mathrm{mode} = \mathbf{P} \, \bm{\sigma}_\mathrm{block} \, \mathbf{P}^\top,
\end{equation}
where
\begin{equation}
    \mathbf{P} =
    \sum_{j=1}^m \left( \mathbf{e}_{2j-1} \mathbf{e}_j^\top + \mathbf{e}_{2j} \mathbf{e}_{m+j}^\top \right),
\end{equation}
\nomenclature{$\mathbf{P}$}{Permutation matrix from block-wise to mode-wise notation \nomrefeqpage}
and \( \mathbf{e}_k \) denotes the standard basis column vector of dimension \( 2m \), with a 1 in position \( k \) and 0 elsewhere.
\nomenclature{$\mathbf{e}_k$}{Canonical (standard) basis column vector with a 1 in the $k$-th position and 0 elsewhere\nomrefeqpage}

\subsection{Quadrature Rotation}
In our simulations of two-mode squeezing generated by a JTWPA, the strongest correlations emerge along the $x$–$x$ quadratures of the two modes. To express the covariance matrices in a cluster-type basis, we apply a local rotation of \( \theta = \pi/4 \) to each mode. This transformation maps the maximal $x$–$x$ correlations into standard $p$–$x$ cluster-type correlations, facilitating comparison with ideal Gaussian cluster states.

This corresponds to a symplectic transformation on the output covariance matrix $\bm{\sigma}$: 
\begin{equation}
    \bm{\sigma_\theta = \textbf{R}_{\theta} \cdot \sigma  \cdot\textbf{R}_{\theta}^{\text{T}}}
\end{equation}
\nomenclature{$\bm{\sigma_\theta}$}{Rotated output covariance matrix \nomrefeqpage}
with $\textbf{R}_{\theta}$ defined in Eq. \ref{eq:rotmat}.
All explicit covariance matrices reported below are expressed in this rotated (mode-wise) basis.

\subsection{Linear Four-Mode Cluster with Distributed Ancillary Coupling: Covariance Matrix}

We now present the symbolic expression for the reduced covariance matrix of a linear four-mode cluster state with six ancillary couplings. This result is obtained by appropriately choosing the adjacency matrix \( \mathbf{A} \) and following the procedure described in Eqs.~\ref{eq:uandv} and \ref{eq:sigmaoutsimb}.

This configuration involves ten modes in total:
\begin{itemize}
    \item Principal modes: \(1\!-\!4\), connected linearly via TMS links \( (1,2), (2,3), (3,4) \).
    \item Ancillary modes: \(5\!-\!10\), connected to the principal chain via TMS links:
    \[
    (1,5),\ (1,6),\ (2,7),\ (3,8),\ (4,9),\ (4,10).
    \]
\end{itemize}

Assuming a uniform squeezing parameter \( r \), we define the following symbolic quantities:
\[
\begin{aligned}
D &= -6 \cosh(|r|) + 3 \cosh(2|r|) + \frac{7}{2}, \\
A &= \left(3 \cosh(|r|) - 2\right) \sinh(|r|), \\
E &= \tfrac{1}{2} \sinh^2(|r|).
\end{aligned}
\]

The reduced covariance matrix for the cluster modes \(1\!-\!4\), in the rotated mode-wise basis, is:
\[
\bm{\sigma}_\mathrm{red}^{(1:4)} =
\left[\begin{matrix}
D & 0 & 0 & A & E & 0 & 0 & 0 \\
0 & D & A & 0 & 0 & E & 0 & 0 \\
0 & A & D & 0 & 0 & A & E & 0 \\
A & 0 & 0 & D & A & 0 & 0 & E \\
E & 0 & 0 & A & D & 0 & 0 & A \\
0 & E & A & 0 & 0 & D & A & 0 \\
0 & 0 & E & 0 & 0 & A & D & 0 \\
0 & 0 & 0 & E & A & 0 & 0 & D
\end{matrix}\right]
\]

\subsection{Cyclic Four-Mode Cluster State with Ancillary Modes: Covariance Matrix}

We now extend the theoretical model of a cyclic four-mode cluster state to include ancillary modes entangled via two-mode squeezing (TMS). This configuration reflects realistic conditions in multi-pump TWPA platforms, where each principal mode is coupled to additional spectral modes through engineered or parasitic squeezing.

We consider a system of twelve modes:
\begin{itemize}
    \item Principal modes: \( 1\!-\!4 \), forming a closed ring via TMS links \( (1,2), (2,3), (3,4), (4,1) \).
    \item Ancillary modes: \( 5\!-\!12 \), coupled to the principal modes as follows:
    \[
    (1,5), (1,6), \quad (2,7), (2,8), \quad (3,9), (3,10), \quad (4,11), (4,12)
    \]
\end{itemize}

Assuming uniform squeezing parameter \( r \), we define the following symbolic coefficients:
\[
\begin{aligned}
D &= -12 \cosh(|r|) + 5 \cosh(2|r|) + \frac{15}{2}, \\
A &= \left(4 \cosh(|r|) - 3\right) \sinh(|r|), \\
E &= \sinh^2(|r|).
\end{aligned}
\]

Then the reduced covariance matrix for modes \(1\!-\!4\), in the mode-wise rotated quadrature basis, is:
\[
\bm{\sigma}_\mathrm{red}^{(1:4)} =
\left[\begin{matrix}
D & 0 & 0 & A & E & 0 & 0 & A \\
0 & D & A & 0 & 0 & E & A & 0 \\
0 & A & D & 0 & 0 & A & E & 0 \\
A & 0 & 0 & D & A & 0 & 0 & E \\
E & 0 & 0 & A & D & 0 & 0 & A \\
0 & E & A & 0 & 0 & D & A & 0 \\
0 & A & E & 0 & 0 & A & D & 0 \\
A & 0 & 0 & E & A & 0 & 0 & D
\end{matrix}\right]
\]

\subsection{Star Four-Mode Cluster State with Ancillary Modes: Covariance Matrix}

We now consider a star-shaped cluster configuration extended to include ancillary modes coupled via two-mode squeezing (TMS) to the outer (leaf) modes. This scenario models a realistic system where the central hub remains clean while each peripheral mode connects to external spectral modes.

The system comprises ten modes in total:
\begin{itemize}
    \item Principal modes: \(1\!-\!4\), arranged such that mode \(1\) is connected to \(2,3,4\).
    \item Ancillary modes: \(5\!-\!10\), with TMS couplings:
    \[
    (2,5), (2,6), \quad (3,7), (3,8), \quad (4,9), (4,10)
    \]
\end{itemize}

Assuming a uniform squeezing parameter \( r \), we define:
\[
\begin{aligned}
D &= -6 \cosh(|r|) + 3 \cosh(2|r|) + \frac{7}{2}, \\
A &= \left(3 \cosh(|r|) - 2\right) \sinh(|r|), \\
E &= \tfrac{1}{2} \sinh^2(|r|).
\end{aligned}
\]

Then, the reduced covariance matrix for the principal cluster modes \(1\!-\!4\), in the rotated mode-wise quadrature basis, is:
\[
\bm{\sigma}_\mathrm{red}^{(1:4)} =
\left[\begin{matrix}
D & 0 & 0 & A & 0 & A & 0 & A \\
0 & D & A & 0 & A & 0 & A & 0 \\
0 & A & D & 0 & E & 0 & E & 0 \\
A & 0 & 0 & D & 0 & E & 0 & E \\
0 & A & E & 0 & D & 0 & E & 0 \\
A & 0 & 0 & E & 0 & D & 0 & E \\
0 & A & E & 0 & E & 0 & D & 0 \\
A & 0 & 0 & E & 0 & E & 0 & D
\end{matrix}\right]
\]

\subsection{Fully Connected Four-Mode Cluster State with Ancillary Modes: Covariance Matrix}

We now consider a fully connected four-mode cluster state where each principal mode is coupled via two-mode squeezing (TMS) to three ancillary modes. This configuration models heavy parasitic interactions common in over-pumped TWPA systems.

We consider a total of sixteen modes:
\begin{itemize}
    \item Principal modes: \(1\!-\!4\), fully connected among themselves via TMS links.
    \item Ancillary modes: \(5\!-\!16\), with each cluster mode coupled to three ancillary modes as follows:
    \[
    \text{Mode 1: } (5,6,7), \quad
    \text{Mode 2: } (8,9,10), \quad
    \text{Mode 3: } (11,12,13), \quad
    \text{Mode 4: } (14,15,16)
    \]
\end{itemize}

Assuming uniform squeezing strength \( r \), we define the following symbolic expressions:
\[
\begin{aligned}
D &= -30 \cosh(|r|) + \frac{21}{2} \cosh(2|r|) + 20, \\
A &= \left(6 \cosh(|r|) - 5\right) \sinh(|r|), \\
E &= \sinh^2(|r|).
\end{aligned}
\]

Then the reduced covariance matrix for the principal modes \(1\!-\!4\), in mode-wise (rotated) block-wise order, is:
\[
\bm{\sigma}_\mathrm{red}^{(1:4)} =
\left[\begin{matrix}
D & 0 & E & A & E & A & E & A \\
0 & D & A & E & A & E & A & E \\
E & A & D & 0 & E & A & E & A \\
A & E & 0 & D & A & E & A & E \\
E & A & E & A & D & 0 & E & A \\
A & E & A & E & 0 & D & A & E \\
E & A & E & A & E & A & D & 0 \\
A & E & A & E & A & E & 0 & D
\end{matrix}\right]
\]

\end{widetext}
\end{document}